\documentclass[sn-mathphys-num]{sn-jnl}% Math and Physical Sciences Numbered Reference Style 
\usepackage{graphicx}%
\usepackage{array} %
\usepackage{multirow}%
\usepackage{amsmath,amssymb,amsfonts}%
\usepackage{amsthm}%
\usepackage{amsfonts}
\usepackage{mathrsfs}%
\usepackage[title]{appendix}%
\usepackage{xcolor}%
\usepackage{textcomp}%
\usepackage{manyfoot}%
\usepackage{booktabs}%
\usepackage{algorithm}%
\usepackage{algorithmicx}%
\usepackage{algpseudocode}%
\usepackage{listings}%

\usepackage{amsmath,amssymb,epsfig,graphicx,color,framed}
\usepackage[english]{babel}
\usepackage{hyperref}
\usepackage{bbm}
\usepackage{amsfonts} 
\usepackage{amsmath}
\usepackage{amssymb}
\usepackage{graphicx}
\usepackage{hyperref}
\usepackage{color}
\usepackage{bm} 
\usepackage{isomath} 
\usepackage{braket}
\usepackage{tikz-cd}
\usepackage{amsmath,amscd}
\usepackage{adjustbox}

\newcommand{\vect}[1]{\vectorsym{#1}} % vectors

\theoremstyle{thmstylethree}%

\begin{document}

\title[Article Title]{Solving wave equation problems on D-Wave quantum annealers}


\author[1]{\fnm{Aigerim} \sur{Bazarkhanova}}\email{aigerim.bazarkhanova@nu.edu.kz}

\author[1]{\fnm{Alejandro } \sur{J. Castro}}\email{alejandro.castilla@nu.edu.kz}

\author*[2]{\fnm{Antonio } \sur{A. Valido}}\email{antonio.valido@ulpgc.es}

\affil[1]{\orgdiv{Department of Mathematics}, \orgname{Nazarbayev University}, \orgaddress{\street{Kabanbay Batyr Ave 53}, \city{Astana}, \country{Kazakhstan}}}

\affil*[2]{\orgdiv{iUNAT, Departamento de Física}, \orgname{Universidad de Las Palmas de Gran Canaria}, \orgaddress{\postcode{35017} \city{Las Palmas de Gran Canaria}, \country{Spain}}}


\abstract{
We solve the one-dimensional Helmholtz equation in several scenarios using the
quantum annealer provided by D-Wave Systems within a pseudospectral framework, where the solution is encoded in an appropriate set of basis functions.
We evaluate the performance of different encoding strategies based on algebraic conditioning and adiabatic considerations, and benchmark their performance
against the classical simulated annealing algorithm. In particular, we analyze
the minimum energy gap, the so-called dynamic range, and the mean squared error to
assess the numerical stability, consistency, and accuracy of the solutions returned
by each strategy. Our work highlights the importance of developing custom embedding techniques ensuring well-conditioned algebraic systems. In particular, we
show that encoding strategies producing full-rank matrices with reduced dynamic ranges
enhance the performance of the quantum annealer
even under polychromatic driving and nontrivial initial conditions. We further
discuss the potential of developing hybrid quantum-classical schemes designed to simultaneously satisfy algebraic conditioning and adiabatic requirements.}

%%================================%%
%% Sample for structured abstract %%
%%================================%%

\keywords{Quantum annealing, D-Wave Systems, Helmholtz equation, Adiabatic Quantum Computation}


%\pacs[MSC Classification]{????}

\maketitle

\section{Introduction}\label{sec1}

Wave equation problems play a fundamental role in various fields of physics and engineering, including electromagnetism \cite{vanbladel20071}, acoustics, and quantum mechanics \cite{arfken20051}. A Fourier analysis is commonly applied by assuming time-harmonic solutions characterized by a given frequency, which eventually leads to the study of the so-called Helmholtz equation \cite{chew19951}. Numerically solving this equation poses significant challenges depending on the boundary conditions and the geometry of the problem, particularly in high-frequency regimes where the wavelength may be much smaller than the characteristic domain size \cite{IHLENBURGt19971}. Traditional computational techniques such as finite difference methods \cite{cohen20021}, finite element methods, or spectral methods \cite{gumerov20031,wright20241} offer effective solutions, but their complexity grows substantially with increasing problem size \cite{klaseboer20221}. This motivates the exploration of alternative approaches, including quantum computing.

Quantum annealing (QA) is a heuristic quantum variational algorithm designed to solve quadratic unconstrained binary optimization (QUBO) problems by encoding their solution in the ground state of a programmed quantum Ising Hamiltonian \cite{yarkoni20221,hauke20201}. This scheme relies on the adiabatic quantum computation paradigm \cite{albash20181}, which aims to evolve a quantum system sufficiently slowly from an initial easy-to-prepare state to a desired final state that encodes the solution of the problem. Unlike gate-based quantum computing \cite{garciaripoll20211,lubasch20251}, which relies on unitary operations and error correction, QA exploits quantum tunneling and superposition to explore multiple solutions simultaneously \cite{pelofske2023,rajak20221}, making it particularly suitable for combinatorial optimization problems \cite{willsch20221}. Interestingly, QA has recently been applied to solve partial differential equations by combining finite element, machine learning, and iterative methods; these studies demonstrate the feasibility of solving differential equations using quantum annealing \cite{conley2023quantum,goes2023burgers,srivastava20191}. However, the extent to which QA provides improved accuracy or efficiency for a broad class of wave equation problems remains an active area of research, with ongoing investigations into its performance relative to high-performance classical heuristic algorithms \cite{yarkoni20221}, such as simulated annealing (SA) \cite{bertsimas1993simulated}.

In the present work, we extend the study of QA combined with spectral methods to solve the Helmholtz equation using the largest commercially available quantum annealers provided by D-Wave Systems, which has pioneered the development of commercial QA and demonstrated applications in logistics, machine learning, and materials science \cite{king2025beyond}. A key challenge lies in encoding the solution of the Helmholtz equation into the D-Wave quantum processing unit (QPU), which imposes constraints on the interactions between problem variables. Indeed, the Helmholtz equation often leads to densely coupled quantum Ising Hamiltonians with intricate energy landscapes, requiring minor embedding techniques to map logical variables onto the sparse quantum hardware. This embedding process introduces additional computational overhead, potentially affecting the performance of quantum solutions. Here, we explore how this issue can be mitigated by implementing improved encoding strategies based on algebraic arguments. In particular, we propose the so-called circulant ansatz, which exploits the cyclic structure of circulant matrices. Another important question concerns the extent to which the evolving quantum Ising Hamiltonian satisfies the adiabatic quantum computation condition characterized by the minimum energy gap \cite{borle2022viable,isermann2021}. In this context, we also investigate the performance of a more sophisticated ansatz within a hybrid quantum--classical approach that maximizes this gap and yields a higher success rate. To evaluate the practical performance of our approach, we benchmark our quantum solver against the SA procedure using several metrics: the mean squared error, which measures the accuracy of the obtained solutions; the dynamic range, borrowed from signal processing, which evaluates the numerical stability of the QUBO formulation; and the minimum energy gap, which assesses consistency with the adiabatic condition. Moreover, as noted in previous studies \cite{schuld2021effect,hunout2025variational}, our results highlight the fundamental role of encoding strategies in achieving higher success rates and lower mean squared errors.

%%%%%%%%%%%%%%%%%%%%%%%%%%%%%%%%%%%%%%%%%%%%%%%%%%%%%%%%%%%%%%%%%%%%%%%%%%%%%%%%%%%%%%%%%%
\section{Helmholtz equation in D-Wave Systems}\label{sec2}
In this section, we first describe the mathematical framework of the Helmholtz
equation and briefly outline the standard formulation used to solve it via the SA
technique on classical computers, as well as through the QUBO method implemented
in QA on D-Wave Systems. We then discuss the main challenges associated with this procedure
and present the proposed ansatz designed to address them.

\subsection{Pseudospectral approximation and simulated annealing}

The one-dimensional Helmholtz  equation reads
\begin{equation}\label{eq:Helmholtz}
    \begin{cases}
    u^{\prime\prime}(x)+\tau^{2}u(x) = F(x), \\
    u(0) = \alpha, u^{\prime}(0) = \beta,
    \end{cases}
\end{equation}
where $u(x)$ is a real, continuous, "well-behaved" scalar function (namely, potential) defined in the interval $x\in[0,2\pi]$, with $\alpha \in \mathbb{R}$ and $\beta\in \mathbb{R}$ representing mixed boundary conditions, and $\tau$ is some constant, which is related to the frequency via some appropriate dispersion relation. For instance, it corresponds to the wave number of the harmonic wave propagation of the electric field in a linear, homogeneous, isotropic media \cite{balanis20121,vanbladel20071}.  We further consider an external electric current or potential source denoted by certain real function 
$F(x)$, so our analysis encompasses a broad class of scenarios described by the scalar wave equation from electromagnetics \cite{chew19951}. It is well-known that the complexity of the solutions of the Helmholtz equation significantly depends on the boundary conditions and the geometry of the problem \cite{arfken20051,cohen20021}.

Beside to be ubiquitous in many areas of physics, mathematics and engineering; we have also addressed Eq. (\ref{eq:Helmholtz}) since its solution admits a closed form expression that allows to assess our procedure \cite{balanis20121}, i.e.  
\begin{equation*}
    u(x)=u_{h}(x)+\frac{1}{\tau^{2}}\int_{0}^{x}\sin{(\tau(x-s))}F(s)ds,
    \label{GSHE}
\end{equation*}
where $u_{h}(x)$ denotes the solution in the homogeneous case, that is
\begin{equation}
    u_{h}(x) = \alpha\cos{(\tau x)} + \dfrac{\beta}{\tau}\sin{(\tau x)},
    \label{HSHE}
\end{equation}
where it clearly manifests the periodic character of the solutions of the wave equation.

From a mathematical point of view, finding an exact solution of   Eq. (\ref{eq:Helmholtz}) constitutes a rather formidable task in most practical situations, especially for high-frequency problems where the wave number is sufficiently large compared to the characteristic size of the domain \cite{feng2009discontinuous}. Here we follow a spectral approach which basically consists of representing the solution of  Eq. (\ref{eq:Helmholtz}) as a finite expansion in a set of certain global basis (or test) functions $\phi_{i}(x)$, that is the pseudospectral approximation \cite{shizgal20151}
\begin{equation}\label{EQPSSF}
u_{N}(x)\approx \sum_{i=1}^{N}\mathrm{w}_{i}\phi_{i}(x),
\end{equation}
where $\mathrm{w}_{i}\in \mathbb{R}$ are unknown coefficients, and $\phi_{i}(x)$ are defined on the interval $x\in[0,2\pi]$ and are usually required to be orthonormal with respect to some appropriate weight function for analytical convenience \cite{shizgal20151}. The choice of these basis functions relies on the specific features of the problem under consideration. In particular, the Fourier series is naturally employed in domains with periodic conditions. In the present work, we shall consider two types of basis functions to implement Eq. (\ref{EQPSSF}): an approximate version of the Fourier series and the so-called circulant ansatz that proves beneficial for convergence and conditioning of the algebraic system constituted by the coefficients $\mathrm{w}_{i}$ (see Secs. \ref{SecTFA} and \ref{SecCA} for further details).

For programming purposes, the pseudospectral approximation is applied along with a spatial discretization of the Helmholtz equation, so we can reformulate the original problem into a grid of discrete points $x_{m}$ with $m = 0,1,\ldots, N-1$, usually called collocation points, i.e.
\begin{equation}\label{Eq:CPoints}
    x_{m} = \frac{2\pi m}{N}.
\end{equation}
We consider collocation points uniformly spaced along the domain $(0, 2\pi)$ 
in order to obtain a regular grid. This choice is commonly adopted to simplify interpolation and differentiation procedures in various numerical methods.
As a result, we obtain an $N+2$-dimensional linear system of equations 
(the two additional equations arise from the boundary conditions) 
with $N$ unknowns, namely $\mathrm{w}_{i}$, which can be compactly expressed as
\begin{equation}\label{eq:HelmholtzSM}
\vect a\bm{\mathrm{w}}=\vect b,
\end{equation}
where $\bm{\mathrm{w}}=(\mathrm{w}_{1},\cdots,\mathrm{w}_{N})^{T}$, and $\vect a$ is a $(N+2)\times N$ real matrix and $\vect b$ is a $N+2$ vector which are fully determined from Eq. (\ref{eq:Helmholtz}) once  Eq. (\ref{EQPSSF}) is substituted. 
Hence, by solving Eq. (\ref{eq:HelmholtzSM}) we equivalently obtain an approximate solution of the original differential equation (\ref{eq:Helmholtz}) evaluated in the grid.

According to algebraic arguments, there is a direct connection between the rank of a matrix and the uniqueness of the solution in optimization problems. In other words, if $\vect{a}$ is not of full rank, this leads to degenerate solutions (multiple solutions with the same cost), which may eventually cause the annealer to return ambiguous results. Hence, one may expect that an appropriate choice of basis functions that ensures full rank can significantly influence the accuracy of the solution to problem~ of 
Eq. (\ref{EQPSSF}).

To guarantee the validity of the proposed solution of Eq.  (\ref{eq:HelmholtzSM}) (or, equivalently Eq. (\ref{EQPSSF})), one may impose that the unknown coefficients $\mathrm{w}_{i}$ minimize the (Euclidean) norm of the error function, i.e. $|u(x)-u_{N}(x)|$. Concretely, by invoking the least-squares method, we further translate the algebraic Eq. (\ref{eq:HelmholtzSM}) into a variational problem with the following objective function (also called cost function),
\begin{align}\label{eq:costfunct}
 \mathcal{C}(\bm{\mathrm{w}})  &= \sum\limits_{m=0}^{N-1} |u_{N}''(x_m) + \tau^{2}u_{N}(x_m) - F(x_{m})|^{2}  + |u_{N}(0) - \alpha|^{2}  + |u_{N}'(0) - \beta|^{2}, \nonumber \\
&=\bm{\mathrm{w}}^{T}\vect q\bm{\mathrm{w}}+\vect l\bm{\mathrm{w}}+C_{0},
\end{align}
where the last two terms ensure that the boundary conditions are fully satisfied, and we have introduced the coefficients,
\begin{equation}\label{eq:costfunctql}
    \vect q=\vect a^{T}\vect a, \qquad 
    \vect l=-2\vect b^{T}\vect a, \qquad C_{0}=\vect b^{T}\vect b.
\end{equation}
Let us observe that Eq.~(\ref{eq:costfunct}) measures how the approximated solution~of Eq.~(\ref{EQPSSF}), with coefficients given by~ Eq.~(\ref{eq:HelmholtzSM}), deviates from the exact solution of the differential Eq.~(\ref{eq:Helmholtz}) at the discrete points~of Eq.~(\ref{Eq:CPoints}). 
In particular, the condition $\mathcal{C}(\bm{\mathrm{w}})=0$ yields the coefficients that exactly reproduce the desired solution at the selected points. 
Classically, several numerical variational procedures can solve the optimization problem~of Eq.~(\ref{eq:costfunct}) with high accuracy.  
To benchmark the performance of the proposed approach, we employ the SA method, which is combined with the Monte Carlo technique through the Metropolis algorithm \cite{hauke20201}. 
This method can be regarded as the classical counterpart of the D-Wave QPU. 
It relies on the observation that the optimization problem~of Eq.~(\ref{eq:costfunct}) is equivalent to finding the lowest-energy state of a classical Ising Hamiltonian, i.e.,
\begin{equation}\label{eq:Ising}
H_{\text{Ising}} =  \sum\limits_{i,j}q_{ij}\mathrm{w}_{i}\mathrm{w}_{j} + \sum\limits_{i}l_{i}\mathrm{w}_{i} + C_{0},
\end{equation}
where the variables $\mathrm{w}_{i}\in \mathbb{R}$ represent classical spin values, and the matrix $\vect q$ and the vector $\vect l$ are given by Eq. (\ref{eq:costfunctql}), whereas $C_{0}$ plays no role for the optimization procedure. 

In Section~\ref{sub1sec2}, we describe how the previous procedure can be accommodated within a heuristic variational quantum algorithm, leading to the QUBO framework and ultimately recasting the approximated solution of the Helmholtz Eq.~(\ref{eq:Helmholtz}) into the ground state of a quantum Ising Hamiltonian experimentally implemented on the D-Wave platform. 
Before proceeding further, it is convenient to describe the basis functions employed in the pseudospectral approximation.
%%%%%%%%%%%%%%%%%%%%%%%%%%%%%%%%%%%%%%%%%%%%%%%%%%%%%%%%%%%%%%%%%%%%%%%%%%%%%%%%%%%%%%%%%%%%%%%%%%%%%%%%%%%
\subsubsection{Truncated Fourier Ansatz}\label{SecTFA}
Attending to the periodic properties of the solution of the Helmholtz equation, one may consider the standard basis functions: that is, $\phi_{n}(x)=\sin(nx)$ or $\phi_{n}(x)=\cos(nx)$. We then express the function (\ref{EQPSSF}) as a finite expansion of the truncated Fourier series, given by
\begin{align}\label{eq:TFS ansatz}
    u_{N}^{TFA}(x) = & \sum\limits_{n=1}^{N/2}\left(\mathrm{w}_{n}^{1}\cos(nx) + \mathrm{w}_{n}^{2}\sin(nx)\right),
\end{align}
where $\mathrm{w}_{n}^{1}$, $\mathrm{w}_{n}^{2}$ are the expansion coefficients - spin variables.  Once the ansatz (\ref{eq:TFS ansatz}) is substituted in the expression of the basis function (\ref{EQPSSF}), we obtain a compact matrix form $\vect a^{TFA},$ which is a dense matrix where almost all entries are nonzero (see App. \ref{AppTFA}).  In the context of QA, this dense structure may significantly increase computational overhead, potentially limiting scalability for larger problem sizes.

Notice that the truncated Fourier series is widely used in solving partial differential equations \cite{garciamolina20221,lubasch20251}, especially when the problem exhibits periodicity, as it transforms differential operators into algebraic equations in Fourier space. It is well-known that this spectral method provides high accuracy compared to traditional finite difference or finite element methods \cite{ng2008comparison}. 

%%%%%%%%%%%%%%%%%%%%%%%%%%%%%%%%%%%%%%%%%%%%%%%%%%%%%%%%%%%%%%%%%%%%%%%%%%%%%%%%%%%%%%%%%%%%%%%%%%%%%%%%%%%
\subsubsection{Circulant Ansatz}\label{SecCA}
Now we introduce the circulant ansatz, it reads \cite{asztalos2024reduced, kalinin2020complexity, wilde1983differential},  
\begin{equation}
    u_{N}^{CA}(x) = \sum\limits_{n=0}^{N-1}\mathrm{w}_{n}h_{n}(x),
    \label{eq:circ ansatz}
\end{equation}
where  the auxiliary functions
\begin{equation}
    h_{n}(x) = \frac{1}{N}\sum\limits_{k=-N/2}^{N/2}\frac{1}{c_{k}}e^{ik(x-x_{n})},
    \nonumber
\end{equation}
$x_{n}$ refers to the collocation points of Eq.~ (\ref{Eq:CPoints}) and
\begin{equation}
    c_{k} = \begin{cases}
         1, \, |k| < N/2,\\
         2, \, |k| = N/2. 
     \end{cases} \nonumber
\end{equation}
Interestingly, this choice retrieves a circulant matrix, namely $\vect a^{CA}$ (see App. \ref{APPCA}). This is a special class of structured matrices where each row is a cyclic shift of the previous one. Indeed, it is well known that their structure makes them particularly suitable for numerical methods, especially in applications such as solving differential equations, image processing, and fast algorithms \cite{gilmour1988circulant}.

%%%%%%%%%%%%%%%%%%%%%%%%%%%%%%%%%%%%%%%%%%%%%%%%%%%%%
\subsection{QUBO and D-Wave Systems}\label{sub1sec2}

As anticipated in the introduction, the QUBO model provides a direct mapping of the optimization problem  of Eq.~(\ref{eq:costfunct}) onto the hardware employed in QA, making them a widely adopted framework in quantum computing. This basically consists on mapping the unknown coefficients $\mathrm{w}_{i}$  into qubit variables $\omega_{i}^{l}\in \{0,1\}$ \cite{criado2022qade}, 
\begin{align}\label{eq:binarization}
\mathrm{w}_{i} =  \sum\limits_{l=0}^{n_{\text{spin}}-1}(-1)^{\alpha(l)}\frac{\omega^{l}_{i}}{2^{l}},
\end{align}
where
\begin{align}
  \alpha(l) = \begin{cases}
    1,  \quad & l=0,\\
    0,  \quad & l\neq 0,
\end{cases}  
\nonumber
\end{align}
with $\omega_{i}^{l}$ being spin variables that we need to determine, and $n_{\text{spin}}$ is the number of spins per variable, such that the solution is expressed as a finite set of bitstrings $\vect \omega\in \{0,1\}^{r}$, with $r=N\times n_{spin}$. After substituting  Eq.(\ref{eq:binarization}) in the linear system of equations (\ref{eq:HelmholtzSM}), the latter can be rewritten as  
\begin{equation}\label{eq:HelmholtzSMB}
\vect A\vect \omega=\vect b
\end{equation}
where $\vect A$ is the block-matrix of size $(N+2)\times r$
\begin{equation}
\vect A = 
\left( \begin{array}{c|c|c|c|c}
-\vect a \quad & 2^{-1}\vect a \quad & 2^{-2}\vect a \quad & \dots \quad & 2^{-({n_{spin}-1})}\vect a
\end{array} \right).\label{eq:capA}
\end{equation}
As similarly in the previous section, by doing  least squares on Eq. (\ref{eq:HelmholtzSMB}) we recast this into an optimization problem, i.e.
\begin{equation}
    \min\limits_{\vect \omega\in \{0,1\}^{r}}\vect \omega^{T} \vect Q\vect \omega+\vect L\vect{\omega}+C_{0}',
    \label{QUBO1}
\end{equation}
where $\vect Q$  and $\vect L$ are fully determined from Eq. (\ref{eq:costfunct}) once the desired ansatz (\ref{EQPSSF}) is substituted (see Sec. \ref{SecTFA} and \ref{SecCA} for further details), that is  
\begin{equation*}
    \vect Q=\vect A^{T}\vect A, \qquad 
    \vect L=-2\vect b^{T}\vect A, \qquad C'_{0}=\vect b^{T}\vect b.
\end{equation*}
Observe that $C'_{0}$ can be neglected in the optimization procedure. Here it is important to note that the choice of an appropriate number of spins $n_{spin}$ is crucial for achieving an effective formulation of the problem. A small value of $n_{spin}$ may be insufficient to accurately represent the problem, whereas a large value $n_{spin}$ significantly increases the size of $\vect Q$ and $\vect L$, thereby leading to greater computational complexity. Therefore, selecting an optimal $n_{spin}$ is essential to balance representational accuracy and computational feasibility. 

Provided the QUBO Eq. (\ref{QUBO1}) where variables $\omega_{i}\in \{0,1\}$, we can reformulate it as an equivalent Ising problem 
\begin{equation}
\min\limits_{ \vect {\tilde{\omega}}\in \{-1,1\}^{r}}\vect {\tilde{\omega}}^{T} \tilde{\vect Q} \vect {\tilde{\omega}}+\tilde{\vect L}\vect {\tilde{\omega}}+\tilde{C_{0}'},
    \label{Ising1}
\end{equation}
with variables $\tilde{\omega}_{i}\in \{-1,1\}$ using the linear transformation $\tilde{\omega}_{i} = 2\omega_{i}-1.$ 
Now we map the solution of Eq. (\ref{Ising1}) into the ground state of certain quantum Ising Hamiltonian $\hat{H}_{Q}$, that is 
\begin{equation}\label{QUBOH}  \hat{H}_{Q}=\sum\limits_{i,j}\tilde{Q}_{ij}\hat{\sigma}_{i}^{(z)}\hat{\sigma}_{j}^{(z)}+\sum\limits_{i}\tilde{L}_{i}\hat{\sigma}_{i}^{(z)} ,
\end{equation}
by  elevating the spin variables $\tilde{\omega}_{i}$ into spin operators, i.e.
\begin{align}
\hat{\sigma}_{i}^{(z)} &= (\otimes_{k=1}^{i-1}
\hat{I})\otimes(\hat{\sigma}^{(z)})\otimes (\otimes_{k=i+1}^{r}
\hat{I}), \nonumber \\
\hat{\sigma}_{i}^{(x)} &= (\otimes_{k=1}^{i-1}
\hat{I})\otimes (\hat{\sigma}^{(x)}) \otimes (\otimes_{k=i+1}^{r}\hat{I}), \nonumber 
\end{align}
where we have introduced the Pauli matrices acting upon the $i$th qubit of a system endowed with $r$ qubits, that is
\begin{equation*}
    \hat{\sigma}^{(z)}  = \left(\begin{array}{cc}
   1  & 0  \\
    0 &  -1
\end{array}\right) \quad \text{and} \quad \hat{\sigma}^{(x)} = \left(\begin{array}{cc}
   0  & 1  \\
    1 & 0
\end{array}\right),
\end{equation*}
with $\hat{I}$ and $\otimes$ representing the $2\times 2$ identity matrix and the tensorial product operation, respectively. As expected, the size of the quantum Ising Hamiltonian in matrix form is $2^{r}$. Hence, the D-Wave Systems determine the ground state by performing a quantum annealing procedure characterized by certain time-dependent Ising Hamiltonian \cite{hauke20201}
\begin{equation}\label{eq:Hamiltonian}
    \hat{H}_{\text{Ising}}(t) = (1-t/T)\hat{H}_{0} + (t/T)\hat{H}_{Q},
\end{equation}
where 
\begin{equation}
    \hat{H}_{0} = \sum\limits_{i}\hat{\sigma}_{i}^{(x)},
    \nonumber
\end{equation}
and $T$ is the maximum time allocated for the evolution. 
The above scheme is based on the adiabatic theorem, which states that if the Hamiltonian parameters vary sufficiently slowly, the system remains close to its instantaneous ground state, provided that it is initially prepared in the ground state of the initial Hamiltonian $\hat{H}_{0}$. 
Consequently, the D-Wave systems evolves toward a state close to the ground state of $\hat{H}_{Q}$, which can then be measured and mapped to the numerical solution of the original problem of Eq.~(\ref{eq:costfunct}) \cite{yarkoni20221}. 
This ensures that the desired solution of the optimization problem of Eq.~(\ref{QUBO1}) is obtained. 
More specifically, the adiabatic theorem requires that $T$ must be sufficiently large compared with the inverse square of the minimum energy spectral gap of~ Eq. (\ref{eq:Hamiltonian}) \cite{albash20181}, that is
\begin{equation}\label{EqADC}
 T = O\left( \frac{1}{g_{\min}^2} \right),
\end{equation}
where minimum energy gap  is \cite{borle2019analyzing} 
\begin{align}\label{DefESG}
    g_{\text{min}} = \min\limits_{t\in [0,T]}|\lambda_{1}(t) - \lambda_{0}(t)|
\end{align}  
with $\lambda_{0}(t)$ and $\lambda_{1}(t)$ be the lowest eigenvalues of $\hat{H}_{\text{Ising}}(t)$. Nonetheless, in practical scenarios, environmental noise, thermal fluctuations, and finite annealing times cause the D-Wave system to operate as a heuristic variational quantum algorithm, so that the above adiabatic condition is not fully satisfied. We explecitly address this issue by computing the energy spectral gap for the Hamiltonian (\ref{eq:Hamiltonian}) in several experiments. In particular, in Sec. \ref{sec3AA} we further explore the influence of the energy spectral gap on the performance of the QA by considering a generic ansatz built on the adiabatic prescription. Here, it is important to note that computing the spectral gap~(\ref{DefESG}) becomes increasingly demanding for large matrices $\vect Q$ and $\vect L$, as the size of the quantum Ising Hamiltonian grows exponentially.

From a classical computational perspective, the task of obtaining the global minimum of  Eq. (\ref{eq:costfunct}) (which immediately determines the coefficients $\mathrm{w}_{i}$ that exactly solve  equation
(\ref{eq:HelmholtzSM}) at the collocation points  of Eq. (\ref{Eq:CPoints})) becomes rather difficult when the energy landscape of the classical Ising Hamiltonian (\ref{eq:Ising}) manifests a large number of local minima \cite{hauke20201}. Simultaneously, from the quantum computational perspective, the restricted connectivity of the D-Wave Systems poses a challenge when mapping the (discretized) Helmholtz equation (\ref{eq:HelmholtzSM}) to the quantum spin system. Since binarization makes matrix $\vect A$ (\ref{eq:HelmholtzSMB}) larger, it follows that the   numerical solution requires long-range couplings in $\vect Q$ and $\vect L$ that may exceed the native connectivity of D-Wave’s architecture, represented by the Chimera or Pegasus connectivity graphs \cite{zbinden2020embedding}. To faithfully accommodate a highly connected  $\vect Q$, the D-Wave quantum processor must employ minor embedding, such that logical qubits representing the problem variables are mapped onto multiple physical qubits, effectively increasing connectivity at the cost of introducing additional noise and overhead \cite{yarkoni20221}. In summary, by selecting a basis-function set that maximizes the rank (ensuring a unique, well-conditioned problem) and controls the sparsity structure (thereby enhancing embedding efficiency and reducing noise) of the matrix $\vect A$, we further contribute to increasing the convergence and the accuracy of the heuristic variational quantum annealing process performed by the D-Wave Systems to solve the Helmholtz equation (\ref{eq:Helmholtz}). 

Our numerical and experimental results support that the above intuition holds in both classical as well as quantum scenarios. Concretely,  in Sec. \ref{sec3} we have extensively compared the performance of the SA and QA schemes for two choices of the basis functions introduced previously: it turns out that the TFA retrieves low-rank and dense $\vect A$ matrices (see Sec. \ref{SecTFA}), whereas the CA provides full-rank, well-structured, banded matrices $\vect A$, reducing embedding complexity (see Sec. \ref{SecCA}). 

%%%%%%%%%%%%%%%%%%%%%%%%%%%%%%%%%%%%%%%%%%%%%%%%%%%%%%%%%%%%%%%%%%%%%%%%%%%%%%%%%%%%%%%%%%%%%%%%%%%%%%%%%%%
\section{Experimental results}\label{sec3}
%%%%%%%%%%%%%%%%%%%%%%%%%%%%%%%%%%%%%%%%%%%%%%%%%%%%%%%%%%%%%%%%%%%%%%%%%%%%%%%%%%%%%%%%%%%%%%%%%%%%%%%%%%%
In this section, we present an extensive comparative analysis of QA performance using the previously introduced ansatz, conducted on the D-Wave QA system, with different numbers $N$ of elements in the approximated solution and different values of $n_{\text{spin}}$, to compute approximate solutions of the Helmholtz equation~(\ref{eq:Helmholtz}) under a range of problem configurations. Specifically, we explore different values of the wave number $\tau$, various combinations of boundary conditions defined by the parameters $\alpha$ and $\beta$, and several choices for the source term $F(x)$. This diverse experimental setup encompasses a wide spectrum of realistic scenarios, thereby providing a robust testbed for assessing the impact of ansatz structure and encoding strategy on the QA optimization process. Observe that the binary nature of $\omega$ ($\omega = \omega^{2}$) allows us to reformulate the QUBO equation~(\ref{QUBO1}) more compactly by adding the linear terms $\vect L$ to the diagonal of the quadratic component $\vect Q$ and thus obtain
\begin{equation}\label{QUBODR}
    \min\limits_{\vect \omega\in \{0,1\}^{r}}\vect \omega^{T} \mathcal{Q}\vect \omega + C_{0}',
\end{equation}
where $\mathcal{Q}$ is referred to QUBO matrix.

We have run the Advantage QPU with the Chimera architecture multiple times, that is $n_{run}=1000$, with identical conditions for each experiment and assess the accuracy of the returned solution by means of the metric mean squared error (MSE), i.e.
\begin{equation}
    \text{MSE} = \frac{1}{n}\sum\limits_{i=1}^{n}\Big(u(y_{i}) - u_{N}(y_{i})\Big)^{2},
    \nonumber
\end{equation}
where $n$ corresponds to the number of evenly spaced points across the interval (0, $2\pi$), $u(y_{i})$ is the actual value of the exact solution retrieved by Eq. (\ref{HSHE}) provided the driving force and the initial conditions, and $u_{N}(x_{i})$ is the value obtained from the QA endowed with the TFA of Eq.~ (\ref{eq:TFS ansatz}) or the CA of Eq.~ (\ref{eq:circ ansatz}). Since we are dealing with an heuristic algorithm it is also convenient to evaluate the performance of QA (endowed with a particular ansatz) by the so-called success rate (SR), which corresponds to the number of times (out of 1000) D-Wave could return exact solution. Simultaneously, we benchmark QA performance against the classical SA algorithm, implemented using the software suite provided by D-Wave, as anticipated.

For our subsequent analysis, we also study the dynamic range (DR) of the QUBO matrix, which can be understood as a measure of the stability of the problem, a concept frequently encountered in imaging and signal processing. This is defined as follows:
\begin{equation}\label{EqDRD}
    \text{DR}(\mathcal{Q}) = \log_{2}\left(\frac{\hat{D}(\mathcal{Q})}{\check{D}(\mathcal{Q})}\right),
\end{equation}
where $\hat{D}(\mathcal{Q})$ and $\check{D}(\mathcal{Q})$ denote the absolute values of the maximum and minimum entries of $\mathcal{Q}$, respectively. In the denominator, only nonzero values are considered \cite{davenport2012pros}.

Finally, we analyze the minimum energy gap associated with the evolving quantum Ising Hamiltonian, as defined  by Eq.~(\ref{DefESG}). Motivated by the fact that this quantity plays a crucial role in adiabatic evolution and directly influences the likelihood of the system remaining in its ground state during annealing, we further study QA performance for an alternative ansatz that relies entirely on maximizing the spectral gap in the last part of this section. Accordingly, we must bear in mind that the rank, DR, and $g_{\min}$ serve as indicators for predicting the performance of QA endowed with the aforementioned ansatz, while SR and MSE assess its actual performance.

%%%%%%%%%%%%%%%%%%%%%%%%%%%%%%%%%%%%%%%%%%%%%%%%%%%%%%%%%%%%%%%%%%%%%%%%%%%%%%%%%%%%%%%%%%%%%%%%%%%%%%%%%%%
\subsection{Algebraic prescription: Truncated Fourier and Circulant ansatzes}\label{sec3TFCA}
This section presents the results for the aforementioned metrics when we consider the ansatz introduced previously (see Secs.  \ref{SecTFA} and \ref{SecCA}) to solve the Helmholtz equation in several scenarios, including inhomogeneous Helmholtz equations with rational and irrational initial conditions, and subject to either monochromatic or polychromatic drivings. Let us recall that determining $g_{min}$ becomes computationally  time-consuming when either $n_{\text{spin}}$ or $N$ increases (recall that the quantum Hamiltonian matrix grows as $2^{N\times n_{\text{spin}}}$), making its evaluation unfeasible in many subsequent experiments.

%%%%%%%%%%%%%%%%%%%%%%%%%%%%%%%%%%%%%%%%%%%%%%%%%%%%%%%%%%%%%%%%%%%%%%%%%%%%%%%%%%%%%%%%%%%%%%%%%%%%%%%%%%%
\subsubsection{Homogeneous Helmholtz equation}\label{subsec1}

To get hands-on with the proposed encoding strategies, we begin with the homogeneous Helmholtz equation with trivial initial conditions, that is 
\begin{equation}\label{eq:exp-1}
    \begin{cases}
    u^{\prime\prime}(x)+ u(x) =0, 0<x<2\pi,\\
    u(0) = \frac{1}{2}, u^{\prime}(0) = 0.
    \end{cases}
\end{equation}
The exact solution of this problem can be readily obtained from Eq. (\ref{HSHE}), i.e. $u(x) = 1/2\cos(x)$. From here follows that setting the parameters $N=2$ and $n_{spin}=2$ is sufficient to completely embed $u(x)$ in the proposed solution (\ref{EQPSSF}) for any of the interested basis functions, so we could expect that both SA and QA (carried on the D-Wave Systems) will be able to exactly solve the problem of Eq.~(\ref{eq:exp-1}). However, in a practical scenario, unless we have some previous information about the latter, the number of basis functions $N$ (in the proposed solution  (\ref{EQPSSF})) or the number of spins $n_{\text{spin}}$ (employed in the QUBO) should eventually be updated, either increased or decreased, through a systematic iterative procedure.

\begin{table}[h]
\caption{The rank, DR and energy minimum gap returned by both ansatz, TFA and CA, are shown for different values of the number of spins, while maintaining fixed the number of basis functions, i.e. $N=2$. The last two columns show a comparison of the SR achieved by both SA and QA.}
\label{tab:tab1}
\begin{tabular*}{\textwidth}{@{\extracolsep\fill}ccccccccccc}
\toprule%
& \multicolumn{2}{@{}c@{}}{$\text{rank}(\vect a)$} & \multicolumn{2}{@{}c@{}}{DR($\mathcal{Q}$)} & \multicolumn{2}{@{}c@{}}{$g_{min}$} & \multicolumn{2}{@{}c@{}}{SR SA (\%) } & \multicolumn{2}{@{}c@{}}{SR QA (\%) } \\\cmidrule{2-3}\cmidrule{4-5}\cmidrule{6-7}\cmidrule{8-9}\cmidrule{10-11}%
$n_{\text{spin}}$ & TFA & CA & TFA  & CA & TFA & CA & TFA & CA & TFA & CA \\
\midrule
2 & 2 & 2 & 3.321 & 4.704 & $19.990\times 10^{-2}$ & $9.994\times 10^{-2}$ & 97.1 & 99.6 & 99.3 & 87.1 \\
3 & 2 & 2 & 5.321 & 6.704 & $5.294\times 10^{-2}$ & $2.572\times 10^{-2}$ & 81.4 & 97.8 & 65.8 & 32.0\\
4 & 2 & 2 & 7.321 & 8.704 & $15.625\times 10^{-3}$ & $7.814\times 10^{-3}$ & 52.4 & 56.1 & 15.8 & 6.6 \\
5 & 2 & 2 & 9.321 & 10.704 & $3.906\times 10^{-3}$ & $1.958\times 10^{-3}$ & 36.8 & 28.0 & 3.1 & 1.2 \\
\botrule
\end{tabular*}
\end{table}

Since QA as well as SA return the exact solution for the explored parameter set, we employ the indicator SR to assess the performance of both ansatzes. Table \ref{tab:tab1} illustrates the results obtained for this metric, together with the rest of the metrics when we change $n_{\text{spin}}$ while maintaining fixed the number of basis functions. Overall, we find out that the TFA demonstrates a slightly better performance for both platforms, SA and QA, in agreement with our previous algebraic arguments, TFA retrieves a marginally lower DR while it exhibits an identical rank to CA. Notice that $g_{min}$ takes slightly larger values for the TFA as well, which turns into a slightly better SR. As expected, one may also observe that the SR of the QA scheme decreases significantly relative to SA as $g_{min}$ approaches zero with increasing $n_{\text{spin}}$. This result stems from the adiabatic theorem, since maintaining adiabaticity requires longer evolution times as $g_{\min}$ decreases according to~Eq.~(\ref{EqADC}), which becomes experimentally more challenging due to thermal fluctuations, bias leakage and other technical issues.

%%%%%%%%%%%%%%%%%%%%%%%%%%%%%%%%%%%%%%%%%%%%%%%%%%%%%%%%%%%%%%%%%%%%%%%%%%%%%%%%%%%%%%%%%%%%%%%%%%%%%%%%%%%
\subsubsection{Inhomogeneous Helmholtz equation with monochromatic driving I}\label{subsec2}

Now we consider the influence of a monochromatic driving in the SA and QA performance, i.e.
\begin{equation}\label{eq:exp-2}
    \begin{cases}
    u^{\prime\prime}(x)+ u(x) = \frac{3}{2}\cos{2x}, 0<x<2\pi,\\
    u(0) = 0, u^{\prime}(0) = 0.
    \end{cases}
\end{equation}
The exact solution is $u(x) = 1/2(\cos{x} - \cos{2x})$, so we shall take at least $N\geq 4$ and $n_{spin}\geq 2.$ Here we are interested in analyzing how the pseudospectral approximation, endowed with TFA and CA, performs in the presence of a driving force. We begin by studying several figures of merit: rank, DR, $g_{min}$, SR for SA and QA as a function of the spin number with fixed $N=4$, the results are shown in Table \ref{tab:tab2}. A quick glance reveals that an increasing $n_{\text{spin}}$  leads to a larger DR, while the minimum gap decreases toward zero, which  results in  a decreasing SR for both SA and QA. This behavior is expected, since the size of the Hamiltonian grows.

\begin{table}[h]
\caption{The rank, DR and energy minimum gap returned by both ansatz, TFA and CA, are shown for different values of the number of spins, while maintaining fixed the number of basis functions, i.e. $N=4$. The last two columns show a comparison of the SR achieved by both SA and QA.}
\label{tab:tab2}
\begin{tabular*}{\textwidth}{@{\extracolsep\fill}ccccccccccc}
\toprule%
& \multicolumn{2}{@{}c@{}}{$\text{rank}(\vect a)$} & \multicolumn{2}{@{}c@{}}{DR($\mathcal{Q}$)} & \multicolumn{2}{@{}c@{}}{$g_{min}$} & \multicolumn{2}{@{}c@{}}{SR SA (\%) } & \multicolumn{2}{@{}c@{}}{SR QA (\%) } \\\cmidrule{2-3}\cmidrule{4-5}\cmidrule{6-7}\cmidrule{8-9}\cmidrule{10-11}%
$n_{\text{spin}}$ & TFA & CA & TFA  & CA & TFA & CA & TFA & CA & TFA & CA \\
\midrule
2 & 3 & 4 & 54.619 & 8.159 & $1.662\times 10^{-3}$ & $1.535\times 10^{-1}$ & 41.1 & 79.11 & 11.2  & 23 \\
3 & 3 & 4 & 56.619 & 10.189 & $2.787\times 10^{-3}$ & $3.882\times 10^{-2}$ & 18.8  & 33 & 1.3 & 2.5\\
\botrule
\end{tabular*}
%\footnotetext{}
\end{table}

From Table \ref{tab:tab2}, one may further observe that the TFA yields a rank-deficient matrix, while the CA produces a full-rank matrix for $n_{\text{spin}}=2,3$. According to the discussion of Sec. \ref{sec2}, this indicates that the former method may lead to an increased number of false positives. On the other hand, a comparison of the two methods shows that DR is consistently larger for the TFA. Notably, the SR obtained with both SA and QA is consistently higher for the CA (i.e., nearly twice that of the TFA), indicating its greater robustness and reliability as model complexity increases.

\begin{table}[h]
\caption{The rank, DR and MSE for SA and QA returned by both ansatz, TFA and CA, are shown for different values of the number of basis functions, while maintaining fixed the number of spins, i.e. $n_{\text{spin}}=2$.}
\label{tab:tab3}
\begin{tabular*}{\textwidth}{@{\extracolsep\fill}ccccccccccccc}
\toprule%
& \multicolumn{2}{@{}c@{}}{$rank(\vect a)$} & \multicolumn{2}{@{}c@{}}{DR($\mathcal{Q})$} & \multicolumn{2}{@{}c@{}}{MSE (SA)} & \multicolumn{2}{@{}c@{}}{MSE (QA)} \\\cmidrule{2-3}\cmidrule{4-5}\cmidrule{6-7}\cmidrule{8-9}
N & TFA & CA & TFA  & CA & TFA & CA  & TFA & CA \\
\midrule
4 & 3 & 4 & 54.619 & 8.189 & 0 & 0 & 0 & 0\\
8 & 7 & 8 & 65.796 & 56.579 & 0  & 0.011 & 0 & 0 \\
10 & 9 & 10 & 65.077 & 57.803 &0  & 0.026 & 0.124 & 0.044\\
18 & 17 & 18 & 69.398 & 59.189 &  0 & 0.065 & 0.5 & 0.241\\
\botrule
\end{tabular*}
\end{table}

Now we fix $n_{\text{spin}}=2$ and investigate how increasing $N$ influences MSE, as well as the rank and the DR. The results are shown in Table \ref{tab:tab3}. Based on the rank analysis, the TFA method exhibits rank deficiency, whereas the second method, CA, consistently yields a full-rank matrix for all values of $N$. Although the DR is relatively large for both ansatzes, it is notably smaller for the CA method. 
Furthermore, the MSE is also lower for the QA endowed with the CA, while for SA equipped with the TFA performs better. These results suggest that maintaining full rank and achieving a relatively smaller DR value contribute to improved model accuracy in QA, as reflected in the decrease of the MSE.

\begin{figure}[h!]
\centering
\includegraphics[width=0.8\textwidth]{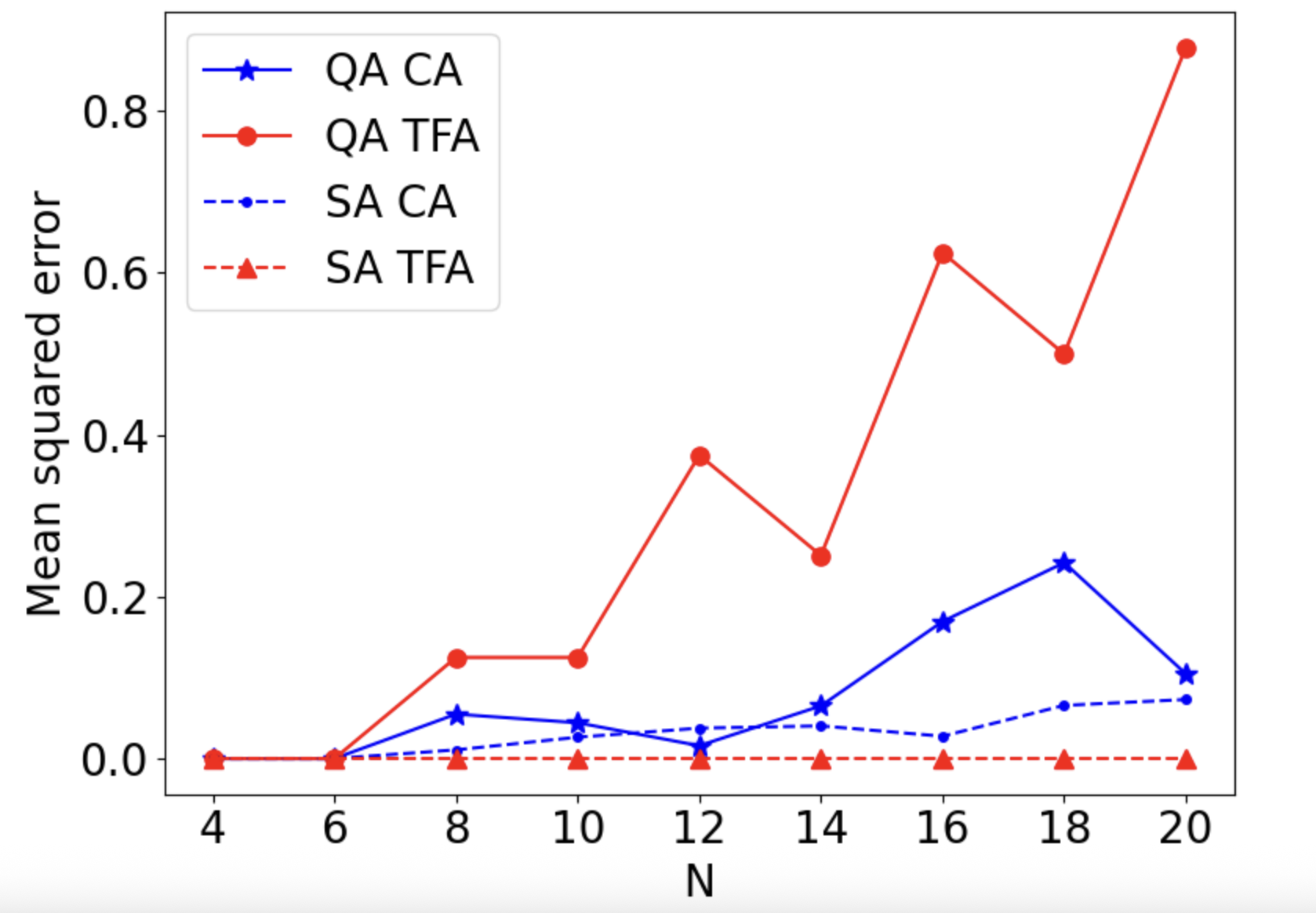} % Adjust the filename and width
\caption{(color online) MSE, retrieved by SA and QA in the scenario of Eq.~(\ref{eq:exp-2}), as a function of the number $N$ of spin functions for a fixed value of the number of spins, i.e. $n_{\text{spin}}=2$.}
\label{fig:fig1}
\end{figure}

Figure \ref{fig:fig1} illustrates the MSE for SA and QA returned by both ansatzes, TFA and CA, as a function of $N$. Since the size of the QUBO matrix grows with $N$, we could only consider $N = 4,\ldots,20$. Across all experiments, the lowest MSE appears for the SA procedure when we use TFA.  However, it is evident that  in the quantum setting, the CA (blue solid line) provides a better approximation than the TFA (red solid line). Overall, increasing $N$ makes it more difficult to reach the ground state. Other practical issues would also affect performance, such as round-off errors during the QUBO/Ising preparation and the limited hardware precision for the Ising coefficients.

\begin{figure}[h!]
    \centering
    
    % -------- Top row --------
    \includegraphics[width=0.32\textwidth]{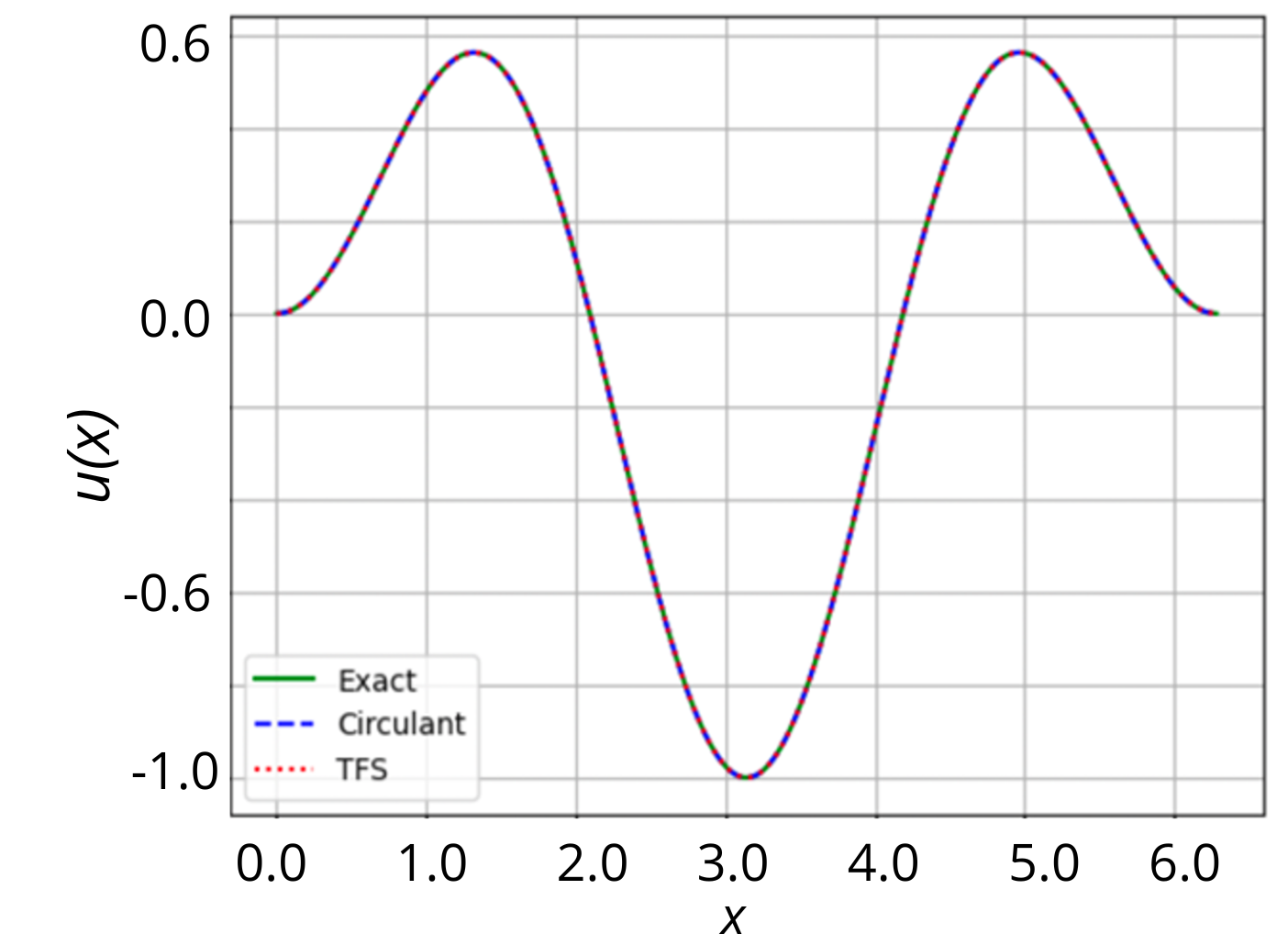}
    \hfill
    \includegraphics[width=0.32\textwidth]{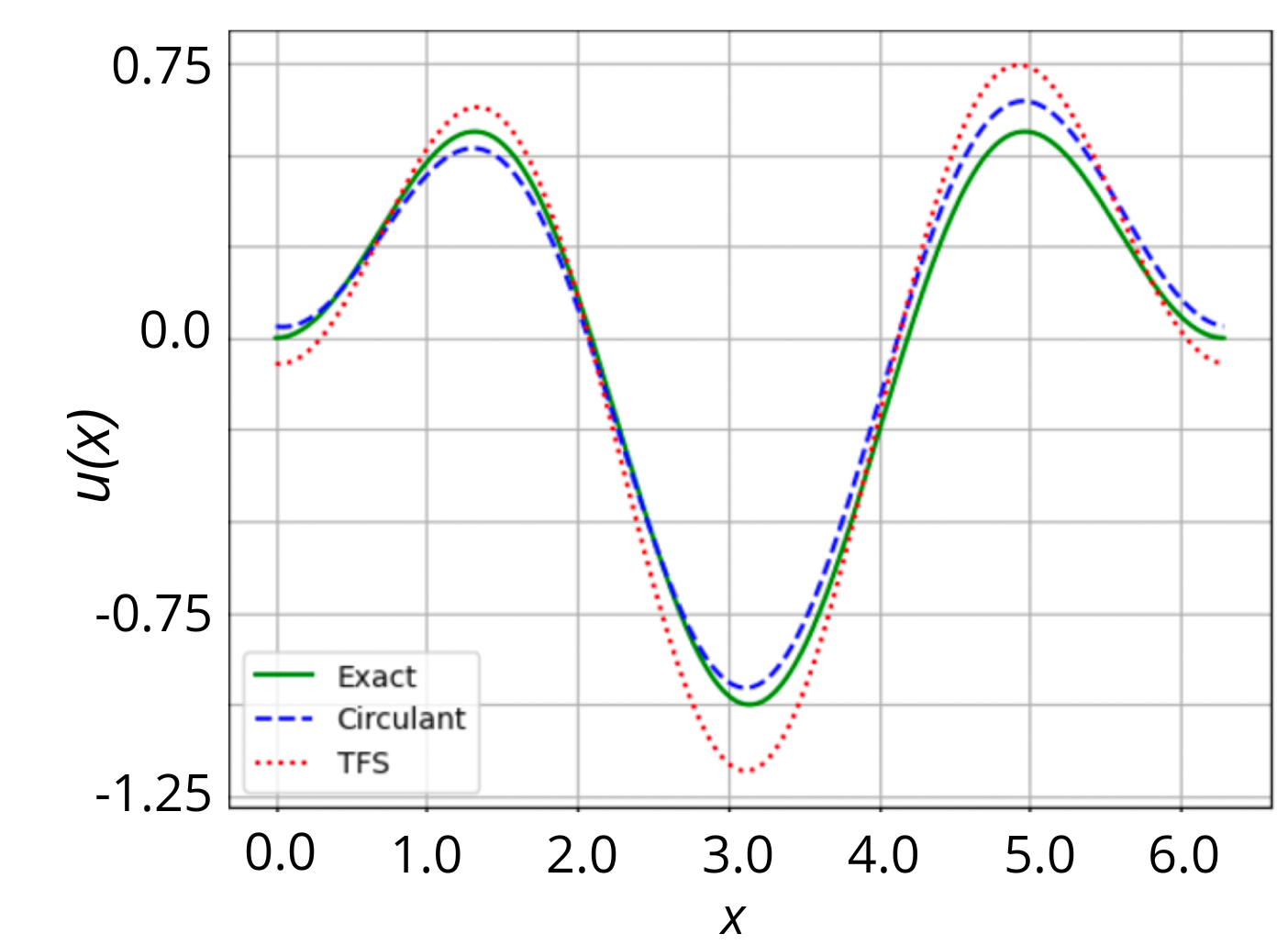}
    \hfill
    \includegraphics[width=0.32\textwidth]{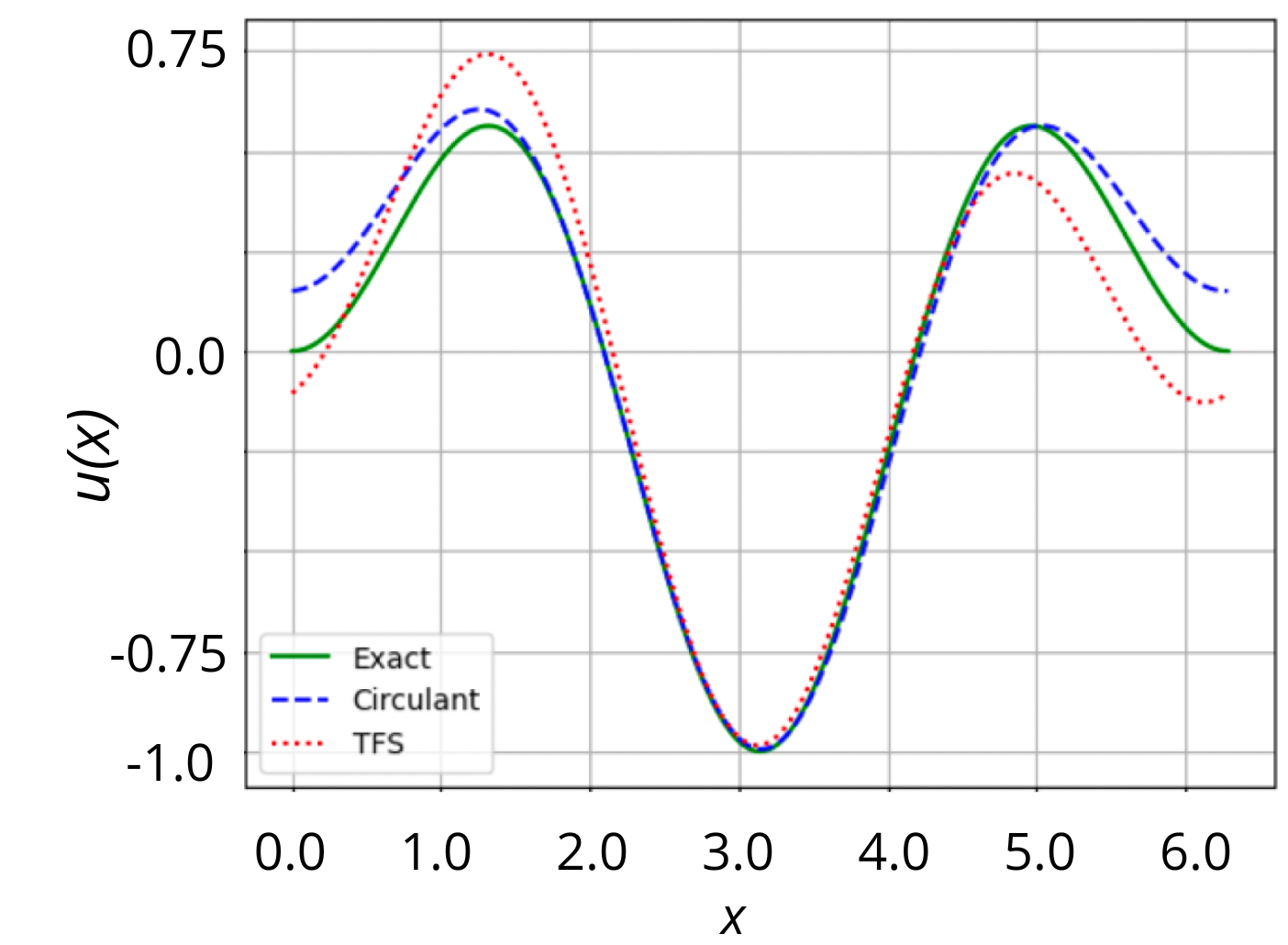}
    
    \vspace{0.4cm}
    
    % -------- Bottom row --------
    \includegraphics[width=0.32\textwidth]{experiment-2,nspin=2.png}
    \hfill
    \includegraphics[width=0.32\textwidth]{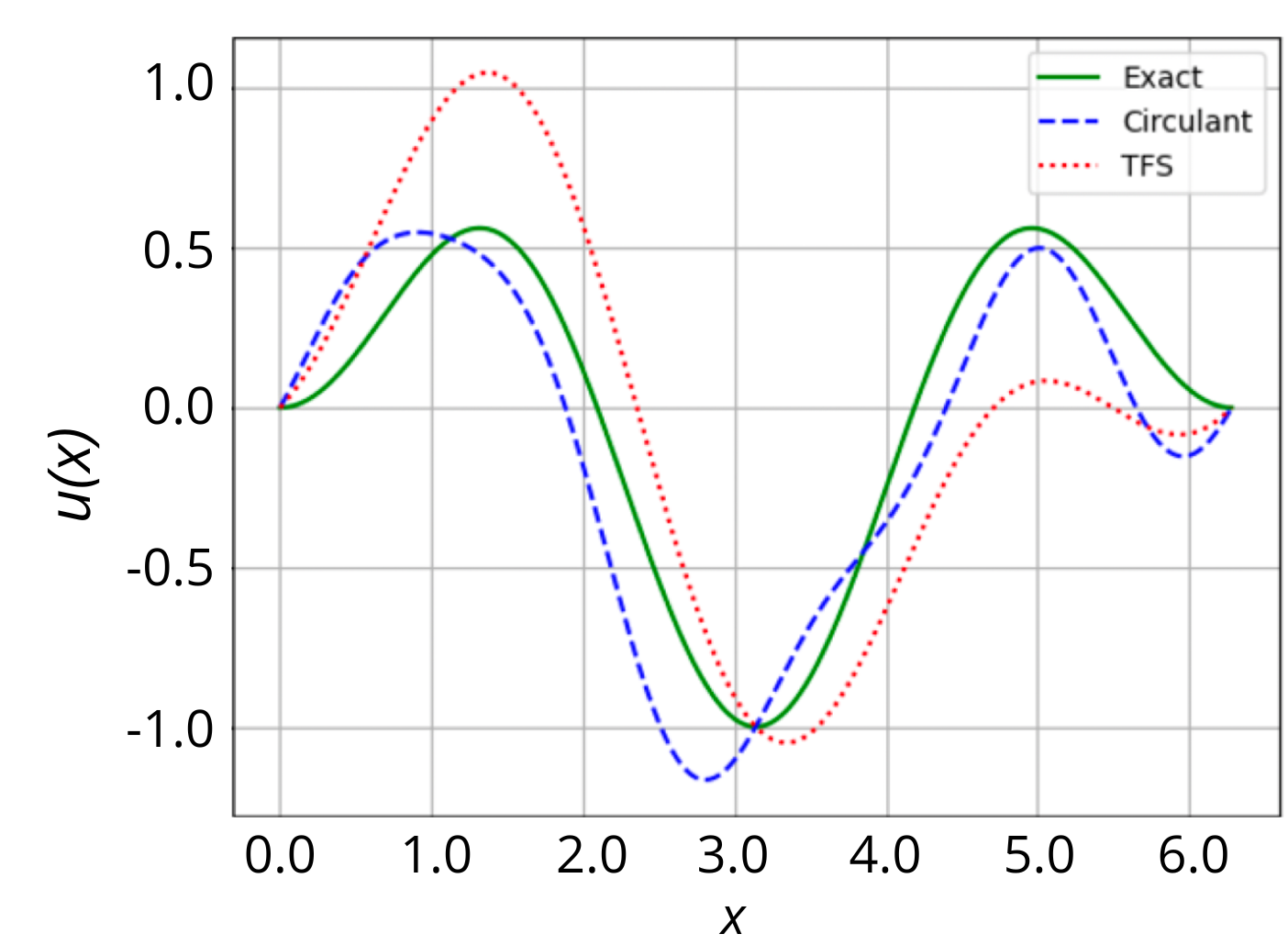}
    \hfill
    \includegraphics[width=0.32\textwidth]{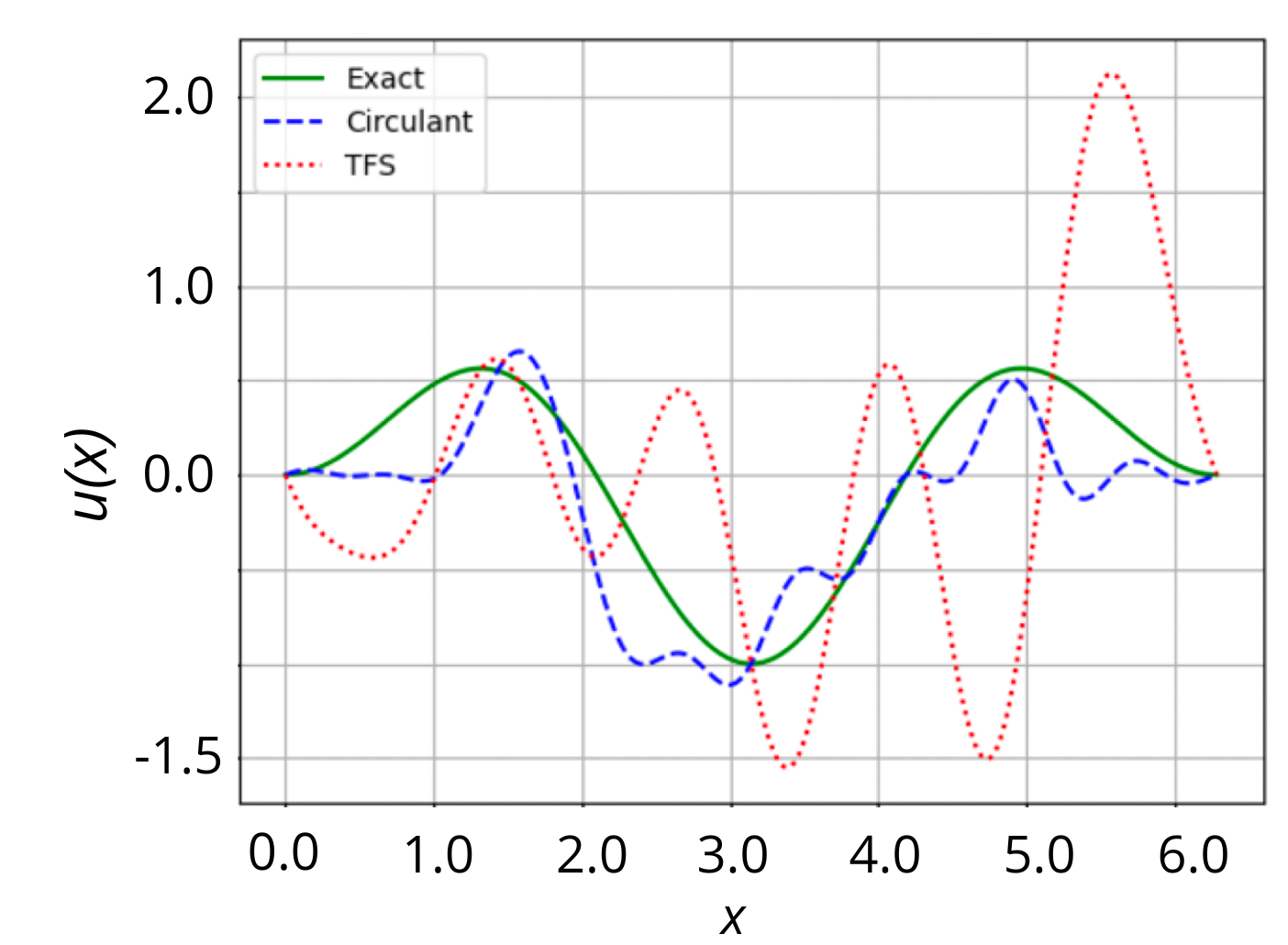}
    
   \caption{(Upper row) Exact and approximated solution returned by QA endowed with TFA and CA for increasing number of spins, $n_{\text{spin}}=2,8,10$ from left to right, and fixed $N=4$. (Lower row) Similarly, exact and approximated solution returned by QA endowed with TFA and CA for increasing number of basis functions, $N=4,10,18$ from left to right, and fixed $n_{\text{spin}}=2$.}
 \label{fig:fig2}
\end{figure}

Finally, Fig. \ref{fig:fig2} shows the approximate solution $u_{N}$, returned by QA endowed with both ansatzes, in comparison with the exact solution. More concretely, the upper panel illustrates this for increasing values of the number of spins, $n_{\text{spin}}=2,8,10$ (from left to right), whereas the lower panel depicts this for increasing values of the number of basis functions, $N=4,10,18$ (from left to right). Clearly, one may observe that across all three cases, the CA consistently demonstrates better performance than the TFA.

%%%%%%%%%%%%%%%%%%%%%%%%%%%%%%%%%%%%%%%%%%%%%%%%%%%%%%%%%%%%%%%%%%%%%%%%%%%%%%%%%%%%%%%%%%%%%%%%%%%%%%%%%%%
\subsubsection{Inhomogeneous Helmholtz equation with monochromatic driving II}\label{subsec3}
In this section, we further investigate the performance of QA when we combine a higher-frequency monochromatic driving with a larger wave number, i.e. $\tau=2$. Concretely, we study the problem
\begin{equation}\label{eq:exp-3}
    \begin{cases}
    u^{\prime\prime}(x)+4u(x) = -6\cos{4x}, 0<x<2\pi,\\
    u(0) = \frac{1}{2}, u^{\prime}(0) = 1,
    \end{cases}
\end{equation}
whose exact solution is  $u(x) = 1/2(\cos{4x} + \sin{2x})$. From problem  of Eq.~(\ref{eq:exp-3}) then follows that the ground state energy would be precisely achieved when $N=8$, and we anticipate that our results confirm this observation. As pointed out, selecting the correct parameter set is important for attaining the lowest energy state: in particular, for this experiment, $n_{\text{spin}}=2$ proves to be sufficient.

\begin{table}[h]
\caption{The rank, DR and MSE for SA and QA returned by both ansatz, TFA and CA, are shown for different values of the number of basis functions, while maintaining fixed the number of spins, i.e. $n_{\text{spin}}=2$.}
\label{tab:tab4}
\begin{tabular*}{\textwidth}{@{\extracolsep\fill}ccccccccccccc}
\toprule%
& \multicolumn{2}{@{}c@{}}{$\text{rank}(\vect a)$} & \multicolumn{2}{@{}c@{}}{DR($\mathcal{Q}$)} & \multicolumn{2}{@{}c@{}}{MSE (SA)} & \multicolumn{2}{@{}c@{}}{MSE (QA)} \\\cmidrule{2-3}\cmidrule{4-5}\cmidrule{6-7}\cmidrule{8-9}%
N & TFA & CA & TFA  & CA & TFA & CA & TFA & CA \\
\midrule
4 & 4 & 4 & 56.619 & 58.064 & 0.495 & 1.251 & 0.125 & 1.054 \\
8 & 7 & 8 & 66.655 & 59.184 & 0.247 & \textbf{0} &0.124  & \textbf{0} \\
10 & 9 & 10 & 64.107 & 58.201 & 0.247 & 0.008 & 0.125 & 0.041\\
18 & 17 & 18 & 69.288 & 59.112 & 0.247  & 0.027 &  0.501 & 0.145\\
\botrule
\end{tabular*}
\end{table}

Similarly to the previous analysis, we fix the number of spins and explore the effects of increasing $N$ on the relevant figures of merits, the results are shown in Table \ref{tab:tab4}. The rank analysis reveals that the TFA method suffers from rank deficiency, whereas the CA consistently maintains a full-rank matrix. While the DR is relatively high for both ansatzes, it is lower in the case of the CA. Clearly, one may observe that the CA yields lower MSE values. These findings are consistent with previous results and, once again, indicate that preserving a full rank matrix $\vect a$ and minimizing the DR are beneficial for enhancing model accuracy, as evidenced by the reduced MSE for QA. Interestingly, by examining the MSE(QA) column in Table \ref{tab:tab4} for both ansatzes, one may further appreciate that the  QA method endowed with the CA retrieves lower values of MSE compared to the TFA, supporting our observation that well-conditioning algebraic structures could help to mitigate noise and control error effects.

\begin{figure}[h!]
\centering
\includegraphics[width=0.7\textwidth]{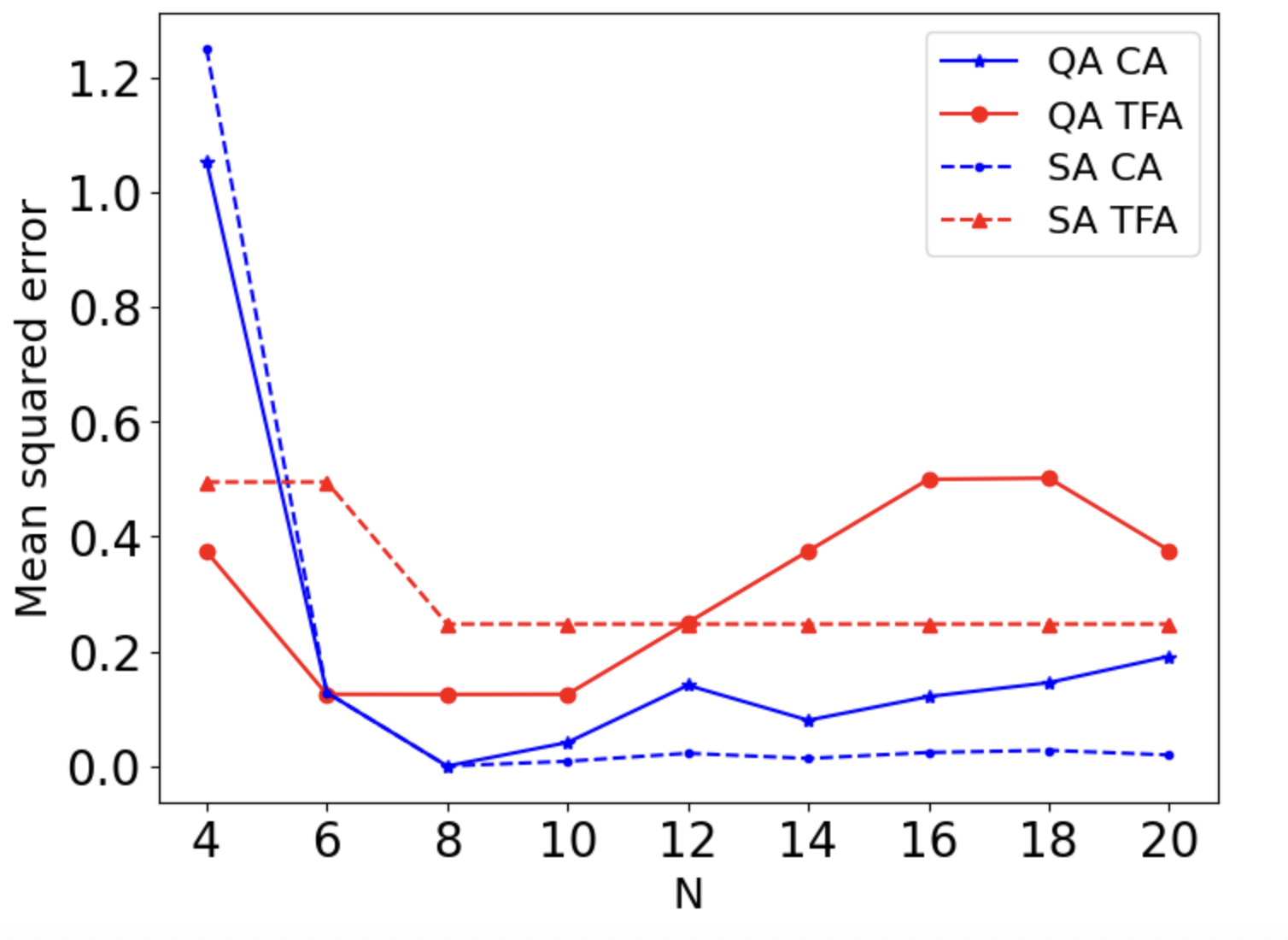} % Adjust the filename and width
\caption{(color online) MSE, retrieved by SA and QA in the scenario  of Eq.~(\ref{eq:exp-3}), as a function of the number $N$ of spin functions for a fixed valued of the number of spins, i.e. $n_{\text{spin}}=2$.}
\label{fig:fig3}
\end{figure}

\begin{figure}[h!]
    \centering
    
    % -------- Top row --------
    \includegraphics[width=0.32\textwidth]{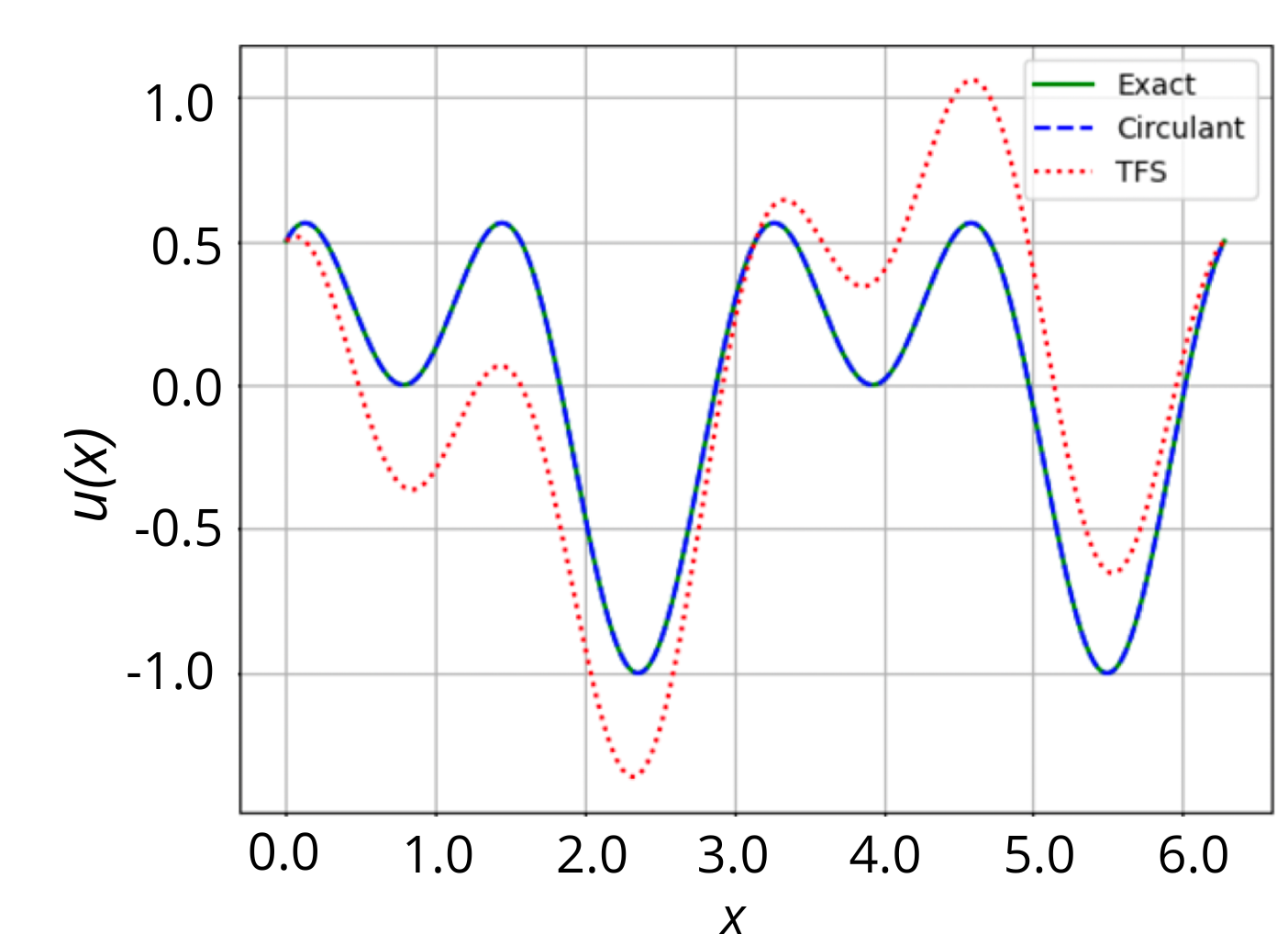}
    \hfill
    \includegraphics[width=0.32\textwidth]{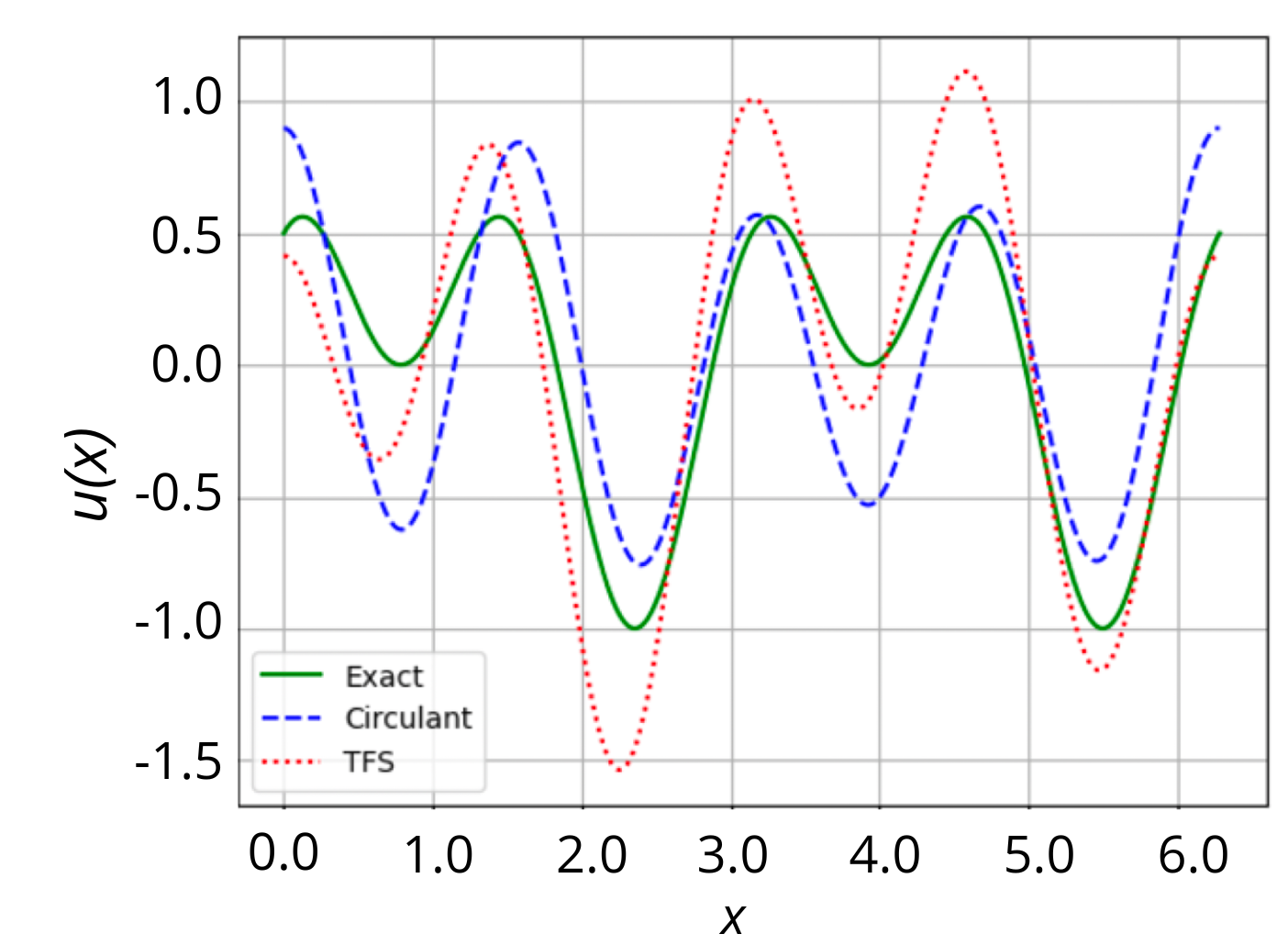}
    \hfill
    \includegraphics[width=0.32\textwidth]{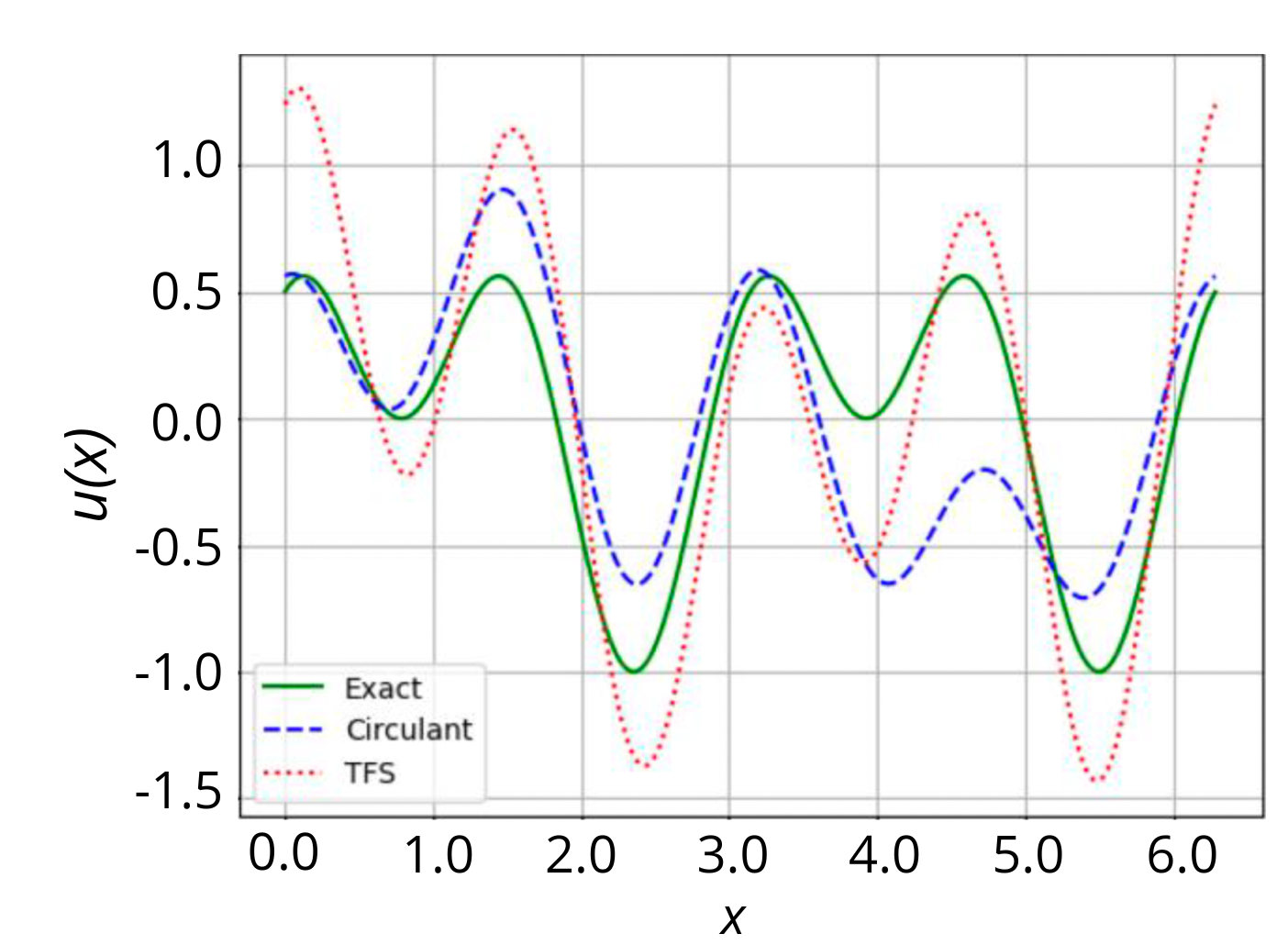}
    
    \vspace{0.4cm}
    
    %-------- Bottom row --------
    \includegraphics[width=0.32\textwidth]{experiment-3,nspin=2.png}
    \hfill
    \includegraphics[width=0.34\textwidth]{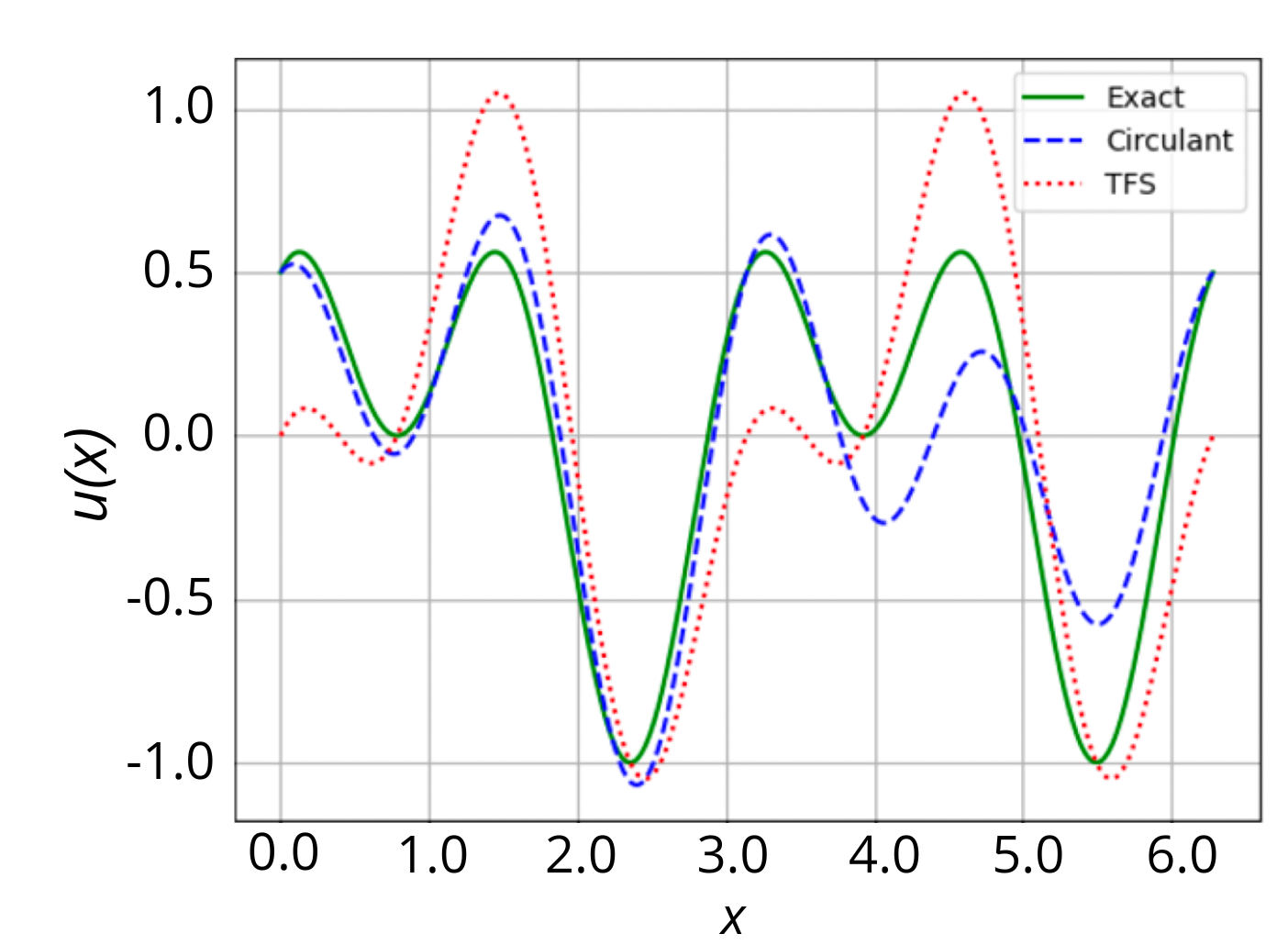}
    \hfill
    \includegraphics[width=0.32\textwidth]{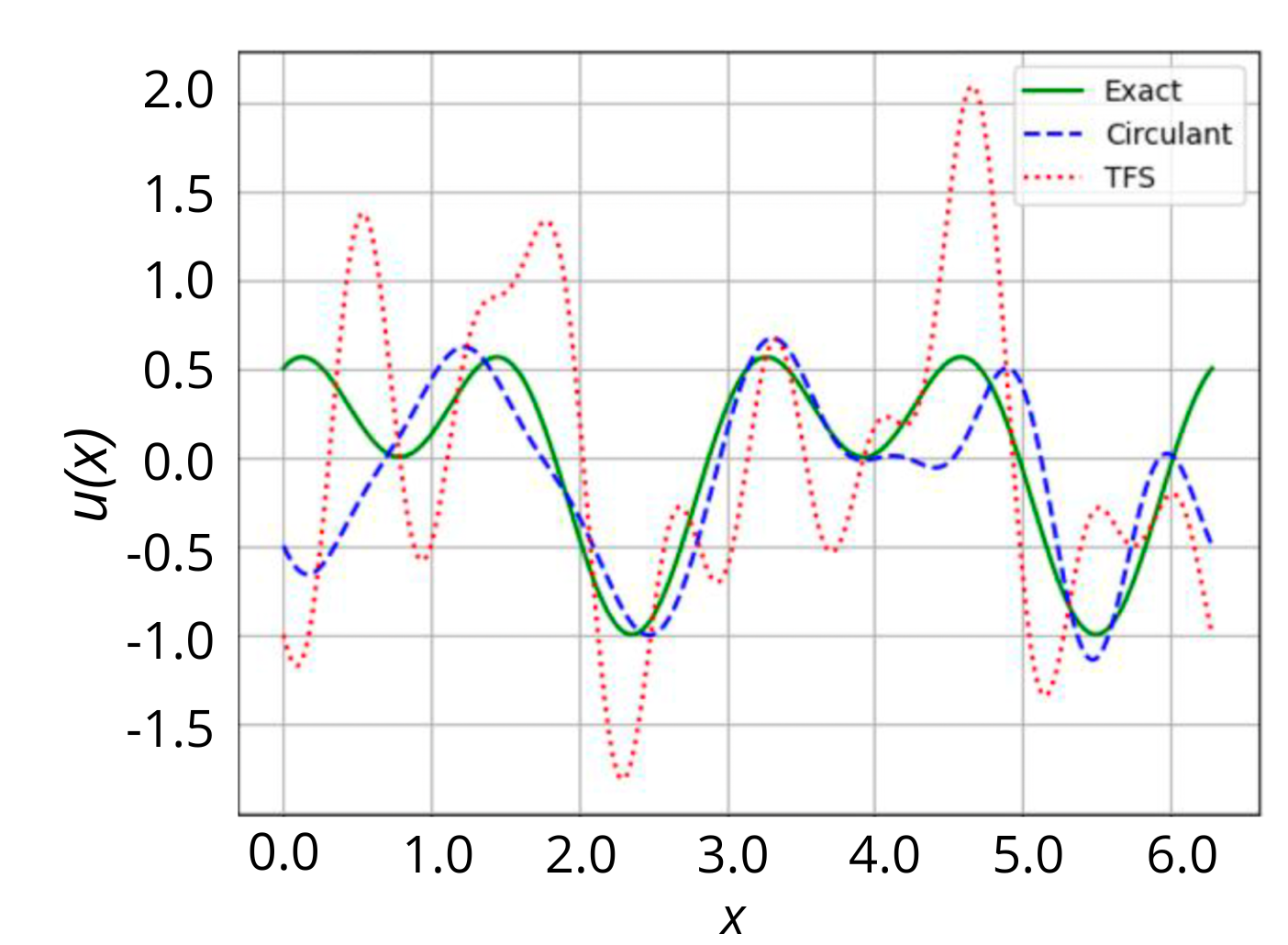}
     \caption{(Upper row) Exact and approximated solution returned by QA endowed with TFA and CA for a fixed number of basis functions (i.e. $N=4$) and an increasing number of spins: $n_{\text{spin}}=2,8,10$ from left to right, respectively. (Lower row) Similarly, exact and approximated solution returned by QA endowed with TFA and CA for a fixed number of spins (i.e. $n_{\text{spin}}=2$) and an increasing number of basis functions: $N=4,10,18$ from left to right, respectively.}
     \label{fig:fig4}
\end{figure}

Interestingly, from Fig. \ref{fig:fig3}, one may see that the CA successfully returns the solution for $N=8$ and $n_{\text{spin}}=2$, whereas the TFA fails to achieve comparable performance for any value of the number of basis functions. These results emphasize that the CA is better suited for more involved initial conditions, in agreement with the discussion of Sec. \ref{sub1sec2}. One may also observe from Fig. \ref{fig:fig4} that the CA method outperforms the TFA method for several values of the number of spins as well (see the left panel in the upper and lower rows). 

We further observe that an increasing number $N$ causes the QA method (combined with either the CA or the TFA) to fail as the MSE increases, while the MSE returned by the SA (endowed with either the CA or the TFA) barely changes. This behavior could be understood by recalling that the embedding task of mapping the problem of Eq.~(\ref{eq:exp-3}) onto quantum hardware may increase crosstalk and control errors as either $N$ or $n_{\text{spin}}$ grow. This eventually makes the adiabatic conditions more difficult to satisfy, thereby degrading the performance of the QA.

%%%%%%%%%%%%%%%%%%%%%%%%%%%%%%%%%%%%%%%%%%%%%%%%%%%%%%%%%%%%%%%%%%%%%%%%%%%%%%%%%%%%%%%%%%%%%%%%%%%%%%%%%%%
\subsubsection{Inhomogeneous Helmholtz equation with polichromatic driving}\label{subsec4}
In this section we pay attention to practical involved situations where the driving is polychromatic, such as
\begin{equation}\label{eq:exp-4}
    \begin{cases}
    u^{\prime\prime}(x)+ u(x) = 2\cos{3x} - \frac{15}{4}\cos{4x} -\frac{3}{4}\sin{2x} + 2\sin{3x}, 0<x<2\pi,\\
    u(0) = -\frac{1}{4}, u^{\prime}(0) = 0.
    \end{cases}
\end{equation}
whose exact solution is $$u(x) = \frac{1}{4}(-\cos(x) - \cos(3x) + \cos(4x) + \sin(x) + \sin(2x) - \sin(3x)).$$ From this expression follows that the solution becomes rather involved for the studied polychromatic driving. 
Overall, the results obtained here are consistent with the discussion of Secs. \ref{sub1sec2} and \ref{subsec3}: the solution retrieved by QA endowed with the CA exhibits higher accuracy, supporting the view that a full rank matrix $\vect a$ and a lower DR can contribute to substantially reduce the MSE in polychromatic scenarios as well. 

Here, we fix the number of spins to $n_{\text{spin}}=3$, since representing the initial condition $\alpha=1/4$ in binary form requires at least three spins. Similarly as before, we study how a growing number $N$ of basis functions influences the performance of rank, DR and MSE in both ansatzes. These results are illustrated in Table \ref{tab:tab5}. As anticipated, the CA produces a full rank matrix $\vect a$  unlike the TFA, and it also returns comparatively lower DR values, as observed in Sec. \ref{subsec3} (although these values may still be relatively high). Notice that the CA again yields the lowest MSE value in both SA and QA.

\begin{table}[h]
\caption{The rank, DR and MSE for SA and QA returned by both ansatz, TFA and CA, are shown for different values of the number of basis functions, while maintaining fixed the number of spins, i.e. $n_{\text{spin}}=3$.}
\label{tab:tab5}
\begin{tabular*}{\textwidth}{@{\extracolsep\fill}ccccccccccccccc}
\toprule%
& \multicolumn{2}{@{}c@{}}{$rank(\vect a)$} & \multicolumn{2}{@{}c@{}}{DR($\mathcal{Q}$)} & \multicolumn{2}{@{}c@{}}{MSE(SA)} & \multicolumn{2}{@{}c@{}}{MSE(QA)} \\\cmidrule{2-3}\cmidrule{4-5}\cmidrule{6-7}\cmidrule{8-9}\cmidrule{9-10}
N & TFA & CA & TFA  & CA & TFA & CA  & TFA & CA \\
\midrule
4 & 3 & 4 & 58.781 & 9.303 &  0.154 & 1.280 & 0.124 & 0.919 \\
8 & 7 & 8 & 68.211 & 60.261 &  0.185 &  0.015 & 0.125 & \textbf{0.054} \\
10 & 9 & 10 & 67.076 & 60.492 &  0.185 & \textbf{0.006} & 0.187 & 0.145\\
18 & 17 & 18 & 71.398 & 61.270 &  0.185 & 0.035 &  0.688 & 0.236\\
\botrule
\end{tabular*}
\end{table}

\begin{figure}[h!]
\centering
\includegraphics[width=0.7\textwidth]{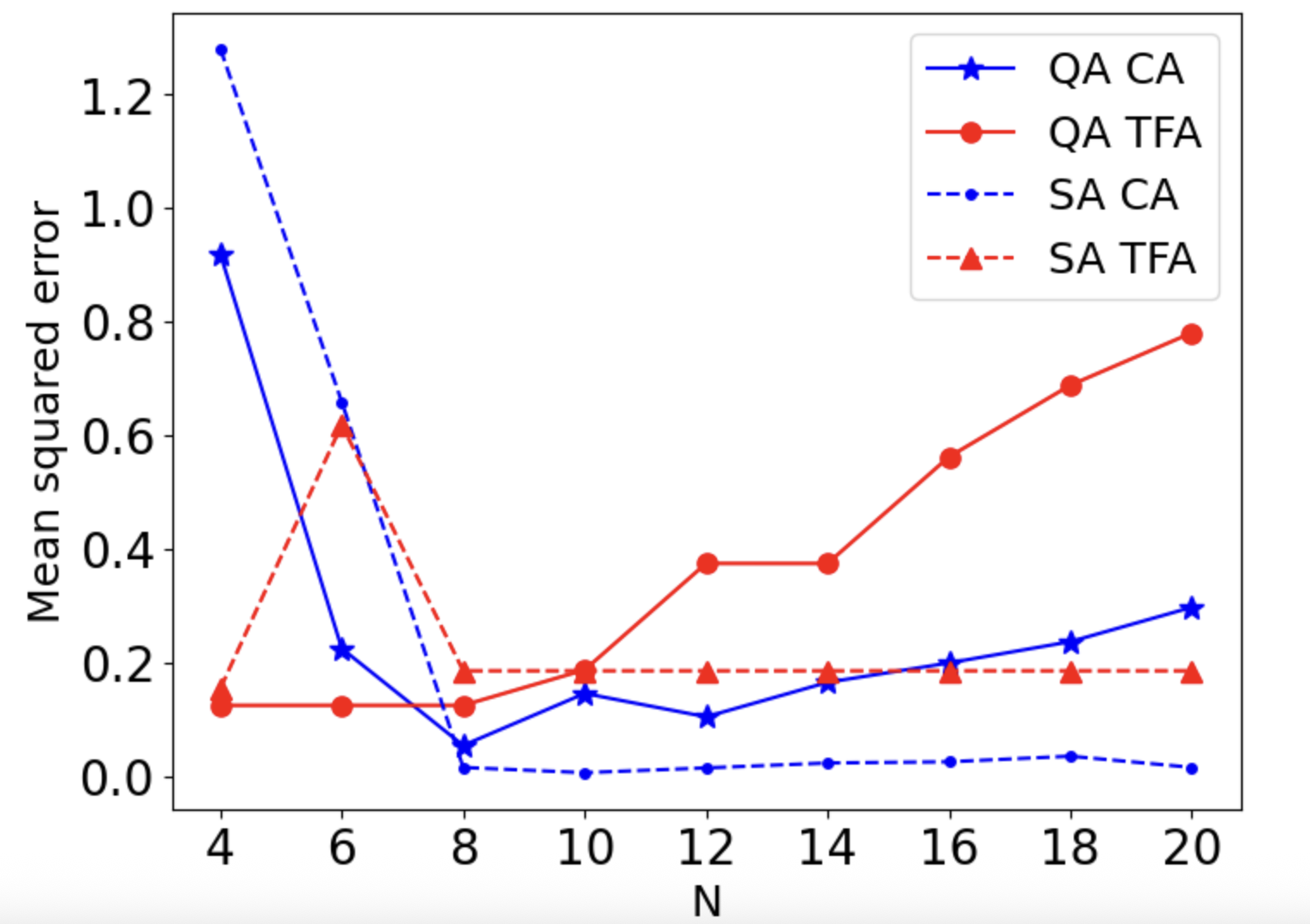} % Adjust the filename and width
\caption{(color online) MSE, retrieved by SA and QA in the scenario of Eq.~(\ref{eq:exp-4}), as a function of the number $N$ of spin functions for a fixed value of the number of spins, i.e. $n_{\text{spin}}=3$.}
\label{fig:fig5}
\end{figure}

Figure \ref{fig:fig5} further illustrates the MSE returned by both ansatzes for a growing number $N$ of basis functions and a fixed number of spins, i.e., $n_{\text{spin}}=3$. From a preliminary Fourier analysis based on  Eq.~\eqref{eq:Helmholtz}, it is expected that both annealing procedures retrieve the ground state for $N=8$. This behavior is observed in Fig. \ref{fig:fig5} where MSE vanishes exactly for $N=8$. Interestingly, we also observe that the SA method, endowed with either the CA or the TFA, converges as $N$ increases, in contrast to  the QA results.

%%%%%%%%%%%%%%%%%%%%%%%%%%%%%%%%%%%%%%%%%%%%%%%%%%%%%%%%%%%%%%%%%%%%%%%%%%%%%%%%%%%%%%%%%%%%%%%%%%%%%%%%%%%
\subsubsection{Homogeneous Helmholtz equation with irrational initial conditions}\label{subsec5}

The purpose of this experiment is to assess the accuracy of the SA and the QA methods when the initial condition $\alpha$ cannot be exactly represented in binary form unless we take an infinitely large number of spins. Concretely, we now study the following problem
\begin{equation}\label{eq:exp-5}
    \begin{cases}
    u^{\prime\prime}(x)+u(x) = 0, 0<x<2\pi,\\
    u(0) = \frac{\sqrt{2}}{2}, u^{\prime}(0) = 0,
    \end{cases}
\end{equation}
whose exact solution is $u(x) = \frac{\sqrt{2}}{2}\cos{(x)}.$

\begin{table}[h]
\caption{The rank, DR and MSE for SA and QA returned by both ansatz, TFA and CA, are shown for different values of the number of spins, while maintaining fixed the number of basis function, i.e. $N=2$.}
\label{tab:tab6}
\begin{tabular*}{\textwidth}{@{\extracolsep\fill}ccccccccccccc}
\toprule%
& \multicolumn{2}{@{}c@{}}{$rank(\vect a)$} & \multicolumn{2}{@{}c@{}}{DR($\mathcal{Q}$)} & \multicolumn{2}{@{}c@{}}{$g_{min}$} & \multicolumn{2}{@{}c@{}}{MSE(SA)} & \multicolumn{2}{@{}c@{}}{MSE(QA)} \\\cmidrule{2-3}\cmidrule{4-5}\cmidrule{6-7}\cmidrule{8-9}\cmidrule{10-11}%
$n_{\text{spin}}$ & TFA & CA & TFA  & CA & TFA & CA  & TFA & CA & TFA & CA\\
\midrule
2 & 2 & 2 & 6.086 & 5.479 & $1.999\times 10^{-1}$ & $1.029\times 10^{-1}$ &  0.021 &  0.021 & 0.022 & 0.022 \\
3 & 2 & 2 & 6.149 & 8.546 & $4.105\times 10^{-2}$ & $2.414\times 10^{-2}$ & 0.001 & 0.001 &  0.001 & 0.001  \\
4 & 2 & 2 & 7.543 & 8.873 & $4.902\times 10^{-3}$ & $4.916\times 10^{-3}$ & 0.001 & 0.001 & 0.001 & 0.001\\
5 & 2 & 2 & 9.543 & 10.873 & $1.509\times 10^{-3}$ & $1.953\times 10^{-3}$ & 0.001 & 0.001 &  0.001 & 0.001 \\
\botrule
\end{tabular*}
\end{table}

From Table \ref{tab:tab6}, one may observe that the rank, the DR, and  the $g_{min}$ exhibit comparable behavior for both the CA and the TFA. In particular, $\vect a$ becomes a full rank matrix, whereas the DR grows as the number of spins increases, as expected. As a consequence, the TFA and CA retrieve similar results for MSE in both methods as well. On the other hand, the minimum energy gap approaches zero as the Hamiltonian size increases, which leads to a corresponding decrease in the SR for both methods (SA and QA). This behavior is consistent with the adiabatic theorem, which states that smaller energy gaps require longer evolution times to maintain the adiabaticity prescription (\ref{EqADC}).

\subsection{Adiabatic prescription: adiabatic ansatz}\label{sec3AA}

Previously, we observed that the ansatzes endowed with higher minimum energy gaps retrieve higher success rates. This observation motivates us to further explore the influence of the energy spectral gap on the choice of the basis function of Eq.~(\ref{EQPSSF}). Concretely, we now build a generic ansatz by appealing to the adiabatic prescription rather than to the previous algebraic arguments: we just require that it maximizes the minimum energy spectral gap (\ref{DefESG}). This alternative ansatz can be expressed as follows

\begin{equation}\label{EqAA}
    u_{N} = \sum\limits_{n=0}^{N-1} \rm{w}_{n} G_{n}(x),
\end{equation}
where $\{G_{0}(x), G_{1}(x), \ldots, G_{N-1}(x)\}$ is a  new set of basis functions, i.e.
\begin{equation}
   G_{n}(x) = \sum\limits_{k=-N/2}^{N/2}g_{n,k}e^{ikx},
   \nonumber
\end{equation}
where $g_{n,k}$ are complex coefficients to be determined. Notice that Eq. (\ref{EqAA}) contains the previously studied ansatzes as particular instances, where  the coefficients $g_{n,k}$ were chosen to retrieve the Fourier series or to satisfy algebraic prescriptions, as occurs in the case of the CA. In contrast, we now impose that the coefficients $g_{n,k}$ maximize the energy spectral gap  (\ref{DefESG}), which can be computed by numerically diagonalizing the quantum Ising Hamiltonian obtained from  Eq. (\ref{EqAA}) following the procedure described in Sec. \ref{sub1sec2}. In particular, this computation was performed using Python numerical libraries in a classical computer. The difficulty of diagonalizing the quantum Ising Hamiltonian (\ref{eq:Hamiltonian}) (recall its size grows as $2^{N\times n_{\text{spin}}}$) makes the optimization procedure a formidable computational task, so we shall focus the attention on the first experiment illustrated in Sec. \ref{subsec1} for fixed values $N=4$ and $n_{\text{spin}}=2$.

Now, using Eq.  (\ref{EqAA}), we compute the ansatz that simultaneously maximizes the energy spectral gap  (\ref{DefESG}) and provides an approximate solution to the problem of Eq.~ (\ref{eq:exp-1}). This will be referred to as the adiabatic ansatz (AA). Table \ref{tab:tab7} shows the results retrieved by both methods, QA and SA, when we employ the latter, as well as the previously studied ansatzes. Interestingly, one may observe that by increasing $g_{\text{min}}$ and ensuring a full rank matrix $\vect a$, we are able to achieve significantly higher values of the SR with the QA method. Concretely, this represents an improvement of approximately $54 \%$ compared to the TFA and $39 \%$ compared to the CA. This contrasts with the SA method, where the AA scheme returns a slightly lower SR compared to the CA. The latter could be understood by paying attention to the DR, this takes a value twice as large for the AA compared to the CA. These results are consistent with our previous observations: higher-rank matrices and narrower dynamic ranges enhance the performance of both annealing approaches. 

The present experiment further suggests that hybrid encoding schemes could be developed to simultaneously provide well-conditioned and stable systems of equations (\ref{eq:HelmholtzSM}) while also maximizing the energy spectral gap. The latter condition ensures that our setup could eventually meet the experimental restrictions imposed by the adiabatic procedure, whereas the former prescription enhances the tractability of the problem by shaping an energy landscape that avoids degenerate regions and numerical instability.

\begin{table}[h]
\caption{The rank, DR and $g_{min}$ returned by TFA, CA and AA are shown for $N=4, n_{spin}=2$. The last two columns show a comparison of the SR achieved by both SA and QA.}
\label{tab:tab7}
\begin{tabular*}{\textwidth}{@{\extracolsep\fill}lcccccccccc}
\toprule%
& \multicolumn{1}{@{}c@{}}{$rank(\vect a)$} & \multicolumn{1}{@{}c@{}}{DR($\mathcal{Q}$)} & \multicolumn{1}{@{}c@{}}{$g_{min}$} & \multicolumn{1}{@{}c@{}}{SR SA (\%) } & \multicolumn{1}{@{}c@{}}{SR QA (\%) } \\ 
\midrule
TFA & 3 & 54.923 & $4.338\times 10^{-4}$ & 44.3 & 8.2 \\
CA & 4 & 7.169 & $1.229\times 10^{-1}$ & 65.9 & 11\\
AA & 4 & 16.19 & $1.332\times 10^{-1}$ & 65.3 & 18 \\
\botrule
\end{tabular*}
\end{table}

\section{Outlook and conclusion}

In this work, we extended the application of QA combined with spectral methods to solve the one-dimensional Helmholtz equation on state-of-the-art quantum hardware provided by D-Wave Systems. Our investigation focused on evaluating the performance of two distinct ansatzes, the TFA and the CA, across different initial conditions and experimental settings, including the number of spectral components $N$ and binary spins $n_{\text{spins}}$. Our study offers a systematic assessment of the strengths and limitations of these encoding strategies for solving the Helmholtz equation, and validates the feasibility of performing QA under current hardware constraints.

Our results consistently demonstrate that the CA, which yields full-rank matrices across various settings, exhibits superior overall performance, including lower mean squared errors and higher success rates. In contrast, the TFA frequently produces rank-deficient matrices, which introduce ambiguity in the solutions and limit its robustness in practical scenarios. The observed trends in the energy spectral gap, as predicted by the adiabatic theorem, align with previous observations and emphasize the importance of suitable algebraic prescriptions in ansatz design for quantum variational heuristics. In particular, a major contribution of our study is the identification of how well-conditioned algebraic properties of the encoded QUBO formulation influence the numerical behavior of QA.

The comparison between quantum and simulated annealing reveals similar qualitative behaviors. In particular, our findings suggest that basis selection and matrix conditioning may help mitigate noise effects and control errors when encoding continuous physical problems into quantum-optimized binary formulations. Collectively, these prescriptions influence the smoothness of the energy landscape and may reduce the likelihood of noisy transitions between near-degenerate solutions, thereby enhancing the robustness of the optimization process in the presence of thermal noise.

Although no industrial application has yet conclusively demonstrated a clear superiority of QA over classical methods, our study contributes to understanding how custom embedding techniques that return full-rank matrices with reduced dynamic range, while preserving D-Wave's architectural constraints, can improve the practical performance of QA in wave-equation problems. Nevertheless, several limitations of the present approach should be acknowledged. 
The scalability of the method is currently constrained by the exponential growth of the quantum Ising Hamiltonian with the number of basis functions $N$ and spins $n_{\text{spin}}$, which increases both preprocessing costs and the difficulty of embedding the resulting QUBO into the quantum hardware. 
In addition, hardware limitations (including restricted qubit connectivity, control errors and thermal fluctuations) may affect the success rate as the problem size increases. 
Furthermore, the present analysis is restricted to a one-dimensional Helmholtz equation with periodic boundary conditions and specific driving terms. 
Extending the framework to higher-dimensional wave equations, more complex geometries, or non-periodic boundary conditions will require further developments in the encoding and embedding strategies. 
Despite these limitations, the results reported here highlight the importance of algebraic conditioning, dynamic-range control, and spectral-gap optimization as guiding principles for designing more robust quantum–classical approaches to wave-equation problems. 
As quantum hardware continues to mature, these insights may contribute to the development of more efficient quantum solvers for wave equations.

\backmatter

\bmhead{Supplementary information}

For supplementary information, the corresponding codes for the experiments are available at the following GitHub repository: \href{https://github.com/aigerimhatake/QA-}{GitHub repository link}.

\bmhead{Acknowledgements}

The authors are grateful to J.J. García-Ripoll for helpful discussions. This work is supported by the Ministry of Science and Higher Education of the Republic of Kazakhstan (Grant No AP19676408).

\begin{appendices}

\section{QUBO Derivation}\label{AppLinSystem}

In this appendix, we start illustrating the derivation of the QUBO Eq. (\ref{QUBO1})  for a general ansatz  (\ref{EQPSSF}) and  in subsections \ref{AppTFA} and \ref{APPCA} we do it  for
the TFA and CA ansatzes.

From the Helmholtz equation (\ref{eq:Helmholtz}) using Eq. (\ref{EQPSSF}) and discretization over the $N$ collocation points of Eq.(\ref{Eq:CPoints}) we get 

$$\begin{cases}
\sum\limits_{i=1}^{N}\mathrm{w}_{i}\phi_{i}^{\prime\prime}(x_{m})+\tau^{2}\sum\limits_{i=1}^{N}\mathrm{w}_{i}\phi_{i}(x_{m}) = F(x_{m}), \quad  m=0,1,\dots, N-1\\
    \sum\limits_{i=1}^{N}\mathrm{w}_{i}\phi_{i}(0) = \alpha\\
    \sum\limits_{i=1}^{N}\mathrm{w}_{i}\phi^{\prime}_{i}(0) = \beta.
\end{cases}
$$

As a result, we get  $N+2$ linear  equations  with unknowns $\bm{\mathrm{w}} = (\mathrm{w}_{1},\mathrm{w}_{2},\dots, \mathrm{w}_{N})\in \mathbb{R}^{N}$, which can be compactly expressed  in matrix form  Eq.(\ref{eq:HelmholtzSM}). Using the binary encoding  of Eq. (\ref{eq:binarization}), we  can obtain the system of equations (\ref{eq:HelmholtzSMB}) for  unknowns $$\vect \omega =(\omega_{1}^{0},\dots,\omega_{N}^{0},\dots, \omega_{1}^{n_{spin}-1},\dots,\omega_{N}^{n_{spin}-1}) \in \{0,1\}^{r}.$$ To reformulate our problem as QUBO we  use the least squares method and get Eq. (\ref{QUBO1}). Next we focus on obtaining the matrix $\vect a$ for each particular case.

\subsection{Linear System of Equations for TFA}\label{AppTFA}

To obtain a compact matrix form for TFA we refer to the ansatz  (\ref{eq:TFS ansatz}) introduced in Section \ref{sec3TFCA}.

Following the method described in the preceding section we obtain
$$\vect a^{TFA}\bm{\mathrm{w}}=\vect b,$$
where 
$$\bm{\mathrm{w}}  = (
\mathrm{w}_{1}^{1}, \mathrm{w}_{2}^{1},\dots, \mathrm{w}_{N/2}^{1},\dots, \mathrm{w}_{1}^{2}, \mathrm{w}_{2}^{2},\dots, \mathrm{w}_{N/2}^{2})\in \mathbb{R}^{N},
$$
and the coefficient matrix is
\begin{equation*}%\label{TFAcoeff}
    \vect a^{TFA} = \left(\begin{array}{cccccccc}
c_{0,1} & \cdots & c_{0,N/2} & s_{0,1} & \cdots & s_{0,N/2} \\
\vdots & \ddots & \vdots & \vdots & \ddots & \vdots   \\
c_{(N-1),1} &   \cdots & c_{(N-1), N/2} & s_{(N-1),1} & \cdots & s_{(N-1),N/2} \\
1 &  \cdots & 1 & 0 & \cdots & 0\\
0 &  \cdots & 0 & 1 & \cdots & N/2
\end{array}\right),
\end{equation*}
for $c_{m,n} = (\tau^{2}-n^{2})\cos{(n x_{m})}$ and $s_{m,n} = (\tau^{2}-n^{2})\sin{(n x_{m})}. $

\subsection{Linear System of Equations for CA}\label{APPCA}

We proceed with the formulation of the circulant  ansatz  (\ref{eq:circ ansatz}) and obtain compact matrix form as
$$\vect a^{CA}\bm{\mathrm{w}}=\vect b,$$
with

$$\vect a^{CA} = 
\left(\begin{array}{ccccccc}
d_{0} & d_{\ell} &  \cdots & d_{(N-2)\ell} & d_{(N-1)\ell}  \\
d_{-\ell} & d_{0}  & d_{\ell} &  \cdots &  d_{(N-2)\ell} \\
\vdots & d_{-\ell} & d_{0}& \ddots & \vdots \\
d_{-(N-2)\ell} & \vdots & \ddots &  \ddots & d_{\ell}\\
d_{-(N-1)\ell} &d_{-(N-2)\ell} & \cdots & d_{-\ell} & d_{0}\\
\hline
e_{0} & e_{\ell} & \cdots & e_{(N-2)\ell} & e_{(N-1)\ell}  \\
f_{0} & f_{\ell}  & \cdots  & f_{(N-2)\ell} & f_{(N-1)\ell}
\end{array}\right) ,
$$
and
$$d_{n\ell} = \sum\limits_{k=-N/2}^{N/2}\frac{(\tau^{2}-k^{2})}{N c_{k}}e^{-ikn\ell}, \quad e_{n\ell} = \sum\limits_{k=-N/2}^{N/2}\frac{1}{N c_{k}}e^{-ikn\ell}, \quad f_{n\ell} = \sum\limits_{k=-N/2}^{N/2}\frac{ik}{N c_{k}}e^{-ikn\ell}$$
for  $n\ell = x_{n}-x_{0} = \frac{2\pi n}{N}$.
In addition, by the periodicity 
\begin{align*}
 e^{-ik(N-j)\ell} = e^{-ikN\ell}e^{-ik(-j\ell)} =  e^{-ikN\frac{2\pi}{N}}e^{-ik(-j\ell)}= e^{-ik(-j\ell)} , 
\end{align*}
we can express 
\begin{equation*}\label{CAcoeff}
    \vect a^{CA} = 
\left(\begin{array}{cccccccccccc}
d_{0} & d_{\ell} & \cdots & d_{-2\ell} & d_{-l}  \\
d_{-\ell} & d_{0}  &  d_{\ell} & \cdots &  d_{-2\ell}  \\
\vdots & d_{-\ell}  &  d_{0} & \ddots & \vdots \\
d_{2\ell} &  \vdots & \ddots & \ddots & d_{\ell} \\
d_{\ell} & d_{2\ell} & \cdots & d_{-\ell} & d_{0} \\
\hline
e_{0} & e_{\ell} & \cdots & e_{(N-2)\ell} & e_{(N-1)\ell}  \\
f_{0} & f_{\ell}  & \cdots  & f_{(N-2)\ell} & f_{(N-1)\ell}
\end{array}\right).
\end{equation*}
Observe that the first block of the matrix provides a circulant structure which is a cyclic shift of the previous one.




\end{appendices}


\bibliography{SWEPQA}

%\input sn-article.bbl
%\printbibliography

%%%%%%%%%%%%%%%%%%%%%%%%%%%%%%%%%%%%%%%%%%%%%%%%%%%%%%%%%%%%%%%%%%%%%%%%%%%%%%%%%%%%%%%%%%%%%%%%%%%%%%%%

\end{document}